\documentclass[usenatbib,useAMS,a4paper]{mn2e}
\usepackage{natbib}
\usepackage{txfonts}
\usepackage{epsfig}
\usepackage{graphicx}
\usepackage{subfigure}

\newcommand{\Msun}{\hbox{$\rm\thinspace M_{\odot}$}}

\newcommand{\Lsun}{\hbox{$\rm\thinspace L_{\odot}$}}

\def\msun{{\rm M_{\odot}}}

\title [Testing radiative cooling approximations]
{ Testing  the accuracy of radiative cooling approximations in SPH simulations}
\author[Daniel R. Wilkins and Cathie J. Clarke]{Daniel R. Wilkins\thanks{E-mail: drw@ast.cam.ac.uk} and Cathie J. Clarke\\Institute of Astronomy, University of Cambridge, Madingley Road, Cambridge CB3 0HA. UK \\
}

\date{Accepted 2011 October 10. Received 2011 September 27; in original form 2011 February 4}

\pagerange{\pageref{firstpage}--\pageref{lastpage}} \pubyear{2011}

\begin{document}
\def\lta{\mathrel{\spose{\lower 3pt\hbox{$\mathchar"218$}}
     \raise 2.0pt\hbox{$\mathchar"13C$}}}
\def\gta{\mathrel{\spose{\lower 3pt\hbox{$\mathchar"218$}}
     \raise 2.0pt\hbox{$\mathchar"13E$}}}
\def\Msun{{\rm M}_\odot}
\def\msun{{\rm M}_\odot}
\def\Rsun{{\rm R}_\odot}
\def\Lsun{{\rm L}_\odot}
\def\19{GRS~1915+105}
\label{firstpage}
\maketitle

\begin{abstract}
Hydrodynamical simulations of star formation have stimulated a need to develop fast and robust algorithms
for evaluating radiative cooling. Here we undertake a critical evaluation of what is currently a popular
method for prescribing cooling in Smoothed Particle Hydrodynamical (SPH) simulations, \textit{i.e.} the polytropic cooling
due originally to Stamatellos \textit{et al}. This method uses the local density and potential to estimate the
column density and optical depth to each particle and then uses these quantities to evaluate
an approximate expression for the net radiative cooling. We evaluate the algorithm by considering
both spherical and disc-like systems with analytic density and temperature
structures. In spherical systems, the total cooling rate computed by the
method is within around $20 \%$ for the astrophysically relevant case of
opacity dominated by ice grains and is correct to within a factor of order unity for a range of opacity laws.
In disc geometry, however, the method
systematically under-estimates the cooling by a large factor
at all heights in the disc. For the self-gravitating disc studied, we
find that the method under-estimates the total cooling rate by a factor
of $\sim 200$.
 This discrepancy may be readily traced to the method's 
systematic 
over-estimate of  the disc column density and optical depth,  since (being
based only  
on the local density and potential) it  does not take
into account the low column density route for photon escape normal to
the disc plane. We note that the discrepancy quoted above applies in
the case that the star's potential is not included in the column
density estimate and that even worse agreement is obtained if the full
(star plus disc) potential is employed. These results raise an obvious
caution about the method's use in disc geometry whenever  an accurate
cooling rate is required, although we note that there are situations
where the discrepancies highlighted above may not significantly
affect the global outcome of simulations. Finally, we draw attention to
our introduction of an  analytic self-gravitating disc structure 
that may be of use in the calibration of future cooling algorithms.
\end{abstract}


\section{Introduction}
Star formation simulations require computationally efficient yet robust algorithms for predicting the thermal evolution of the gas during collapse and fragmentation. A cheap and widely used approach (particularly in SPH simulations) is to simply adopt a barotropic equation of state, whose form is motivated by the results of spherical collapse calculations that include a treatment of radiative transfer \citep{masunaga,larson}. This prescription correctly captures three broad phases of the thermal evolution of the gas, \textit{i.e.} an isothermal stage at low density followed by successive stages where the gas heats up in response to increasing density. However, by associating gas of a particular density with a corresponding temperature it naturally ignores a variety of effects that in practice will affect the thermal state of the gas (\textit{e.g.} the geometry, the temperature of the surrounding gas and the local radiation field).

At the other end of the spectrum of complexity, there is a long history of grid based hydrodynamic simulations that follow the thermal evolution of the gas by including energy transfer between the gas and radiation field. Such calculations vary in the level of sophistication with which they model the radiation field, ranging from full frequency dependent radiative transfer (\textit{e.g.} \citet{wolfire,yorke}) to the widely employed and cheaper expedient of grey flux limited diffusion (\textit{e.g.} \citet{bodenheimer,boley,boss-08,cai,krumholz}). More recently, several groups have implemented the latter approach within SPH simulations \citep{wb04,wb06,whitehouse,mayer} and applied this both to large scale star formation simulations \citep{bate} and to the fragmentation of self-gravitating discs around young stars \citep{meru-bate}.
There is however a substantial computational overhead associated with such calculations and this can impose undesirable  limitations on the resolution 
achievable

An intermediate approach is represented by the algorithm proposed by \citet{stamatellos} and developed by \citet{forgan}. In its original form, this algorithm  (which we henceforth refer to as `polytropic cooling') computed  a local cooling rate based {\it purely on local variables}.\footnote{Note that there is an obvious computational advantage in using quantities that are available for each particle and are pre-computed in the code. The addtional motivation for using the 
potential is that it contains information about the wider environment that
may relate to 
the column density of over-lying gas} Specifically, it equates the cooling rate per unit mass with the local flux ($F$) divided by the column density ($\Sigma$) of material overlying a particular SPH particle. In the case of optically thin gas that is not subject to external irradiation, this reduces to a cooling rate per unit mass that is simply $ \sim \kappa \sigma T^4$ (where $\kappa$ is the opacity, $\sigma$ the Stefan Boltzmann constant and $T$ the temperature). This result is correct in the case of  non-irradiated optically thin gas in local thermodynamic equilibrium. In the case of optically thick gas in local thermdynamic equilibrium, the correct expression for the cooling rate per unit mass is $\nabla \cdot F/\rho$ where $\rho$ is the local density. As noted above, the polytropic cooling prescription  instead sets this to $F/\Sigma$, which is not exactly the same but is likely to be of the same order provided that all quantities (including $F$) vary smoothly over a similar scale length.

In order to calculate $F/\Sigma$ from local variables, the polytropic cooling method makes a further assumption, that it is possible to estimate $\Sigma$ from the local value of the density and potential. In order to achieve this, it is assumed that the relationship between potential ($\Psi$), density  (known for each particle in the SPH code) and $\Sigma$ (to be estimated) can be represented by a mass weighted average over an equivalent $n=2$ polytropic sphere in hydrostatic equilibrium.\footnote{The structure of this equivalent sphere is calculated only from the local density and potential and would therefore not in general be in hydrostatic equilibrium given the actual temperature of the particle.} This estimator therefore purely uses the $n=2$ polytrope as a smoothly varying spherical distribution from which one can conveniently make a rough estimate of the relationship between $\rho$, $\Psi$ and $\Sigma$. It  does not require that the gas in the simulation is in hydrostatic equilibrium. In fact, the estimated relationship between $\rho$, $\Psi$ and $\Sigma$ (see Appendix) is very similar if one instead bases the mapping on a polytrope of different index, $n$, with the coefficient of $\Sigma$ (Equation \ref{SigmaBar.equ}) varying by around $10 \%$ when $n$ is varied over the range $0-5$ \citep{stamatellos}. Evidently, it is the smoothness of the distribution, rather than the specific profile,  that is the important property in this mapping procedure. Note that although the rationale for this approximation is based on the structure of smooth spherical distributions (\textit{i.e.} where all quantities vary over a scale length of order the radius within the sphere) it has also been widely applied in simulations in which discs form and, as such, it is implicitly assumed that this mapping works in planar geometry.

There are several variants of  the above algorithm in use. For example, the cooling prescription also usually contains a term that prevents cooling below a background temperature (see Appendix). More recently, a hybrid method has been developed which combines the polytropic
cooling  method with flux limited diffusion \citep{forgan}. This development follows from the recognition that a possible drawback of the original method is that it always implies cooling of gas (provided that it is hotter than the background temperature), whereas in optically thick regions, the sign of $\nabla \cdot F$ might actually imply that a fluid element is heated by adjoining hotter regions. Flux limited diffusion treats this regime correctly but does not provide a good estimate of the cooling rate per unit mass in regions of low optical depth on the periphery of a condensation. It has been suggested that the advantages of both methods can be combined by simply adding the cooling rates from  flux limited diffusion and from polytropic cooling. Clearly this can only be an improvement if each component  of  the cooling prescription makes a small contribution to the total cooling in the region in which this component is regarded as unreliable. This, however, has not been demonstrated to date.

 The development of such approximate cooling prescriptions represents an important opportunity to improve the verisimilitude of star formation simulations at a computational expense that is minimal compared with a full treatment of radiative transfer in the gas. It is obviously important that such prescriptions are rigorously tested. To date, the tests have consisted of modelling the collapse of a $1 M_\odot$ spherical cloud \citep{masunaga}, the simulation of
the collapse of a rotating cloud \citep{boss_bodenheimer-79} and of a
self-gravitating disc \citep{hubeny} 
and the thermal relaxation of a static spherical cloud \citep{spiegel}. In each case, the simulation results have been compared either with previous radiative transfer calculations or else with analytic results.  Of these tests, only the
latter, known as the Spiegel test,  provides a direct measure of the cooling rates. This  test (which involves the introduction of sinusoidal temperature variations at a range of wavelengths) demonstrates that
polytropic cooling  provides a good measure of the cooling rate in the optically thin limit (as indeed it should, given that
the cooling rate per unit mass in this limit is independent of the accuracy of the optical depth/column density estimate).  However it cannot reproduce the results of this test  (for arbitrary temperatute perturbation waevelengths) 
in the optically thick limit because in this case the cooling rate should depend on the wavelength, as this controls the rate at which radiation diffuses between hotter and colder regions. The polytropic cooling method, however, gives cooling
rates that are independent of the wavelength, since this rate is computed only
using information about the local temperature and the estimated optical depth to the exterior, rather than on any temperature variations within the structure.
The good match to the Spiegel test results in the optically thick limit found
by 
\citet{stamatellos} in fact does not hold if one varies the
wavelength from their adopted value.    
Note that  this drawback has been addressed by the hybrid method of \citet{forgan}
(which correctly reproduces the results of the Spiegel test in the optically thick limit) since it is able to diffuse heat between hotter and
cooler regions.

The other two tests that have been conducted  can instead be regarded as `macroscopic' tests inasmuch as they look at the evolution of the entire system. A natural question that arises  with such tests is whether their outcomes are as sensitively dependent on   a correct treatment of the thermal physics as are the physical situations that will be modelled with the algorithm. This is a hard question to answer unless one has also carried out `microscopic' tests --- in other words, an evaluation of the types of situation in which polytropic cooling  does (or does not) provide a good estimator of the cooling rate. Once understanding of this question has been gained, one can then examine the physical situations that are encountered in real star formation simulations and decide whether or not the algorithm is likely to be providing a reliable measure of the cooling rate.

With this in mind, we here report on a suite of tests using simple analytic
structures in spherical and axisymmetric geometries. We first (\S\ref{coldensity.sec}) compare the predicted values of column density ($\Sigma$) 
and optical depth ($\tau$) provided by the polytropic
cooling  method with their true values obtained by integration of the 
analytic density structures. We then proceed (\S\ref{cooling.sec}) to compare the polytropic estimate of the cooling rate per unit mass (\textit{i.e.} $F_s/\Sigma$ where $F_s$ is the polytropic estimate of the radiative flux) with 
an equivalent quantity derived using true values of the column density and
(in the optically thick limit) compare each of these quantities with
what we term the `true' cooling rate; this being related to the divergence of the 
radiative flux computed in the radiative diffusion approximation. Section 4 summarises our conclusions.  

We caution at the outset that such a microscopic examination of the performance of the algorithm is designed to expose its weaknesses and may  expose weaknesses  which actually turn out to be irrelevant in real simuations. If, for the sake of argument, we find a regime where the method produces cooling rates that are many orders of magnitude different from the true values, this may not actually matter in practice. This is the case if, for example, both true and approximate values lie firmly in the regime where the gas is behaving adiabatically or in a regime where the gas temperature is controlled by the ambient cloud temperature. Here we simply present our results  in order that those who use this method in hydrodynamical simulations can assess the significance {\it in practice} of the discrepancies that we highlight here.

\section{Comparison of column density and optical depth estimates}
\label{coldensity.sec}

\subsection{The application of the polytropic method to spherical hydrostatic structures}

We set up a set of simple model density and temperature distributions corresponding to various polytropes in hydrostatic equilibrium. Hydrostatic structures are of interest when testing the radiative cooling approximation since these systems will, once formed, tend to exist for prolonged periods in hydrodynamical simulations and their thermal evolution in a given state will be important as well as their dynamical evolution.

Given an assumed dependence of the Rosseland mean opacity on density and temperature, we can readily compute the column density ($\Sigma$) and optical depth ($\tau$) along a radial path from any given point to infinity. We then compare these quantities t the column densities and optical depths predicted by the polytropic cooling  method ($\Sigma_{poly}$ and $\tau_{poly}$; see Appendix).
We emphasise that $\tau_{poly}$ is {\it not} simply the product of the local opacity and the estimated column density but also contains a (opacity law dependent) factor that estimates the ratio of local opacity to the path averaged opacity within the polytropic smoothing formulation; see \citet{stamatellos}
and Appendix \ref{kappabar.sec}. This aspect of the polytropic cooling formulation is designed to improve the optical depth estimate in cases where the opacity is a strongly varying function of temperature (and hence position) as in the so-called `opacity gap' in protostellar discs.

\begin{figure}
\begin{center}
\includegraphics[width=8cm]{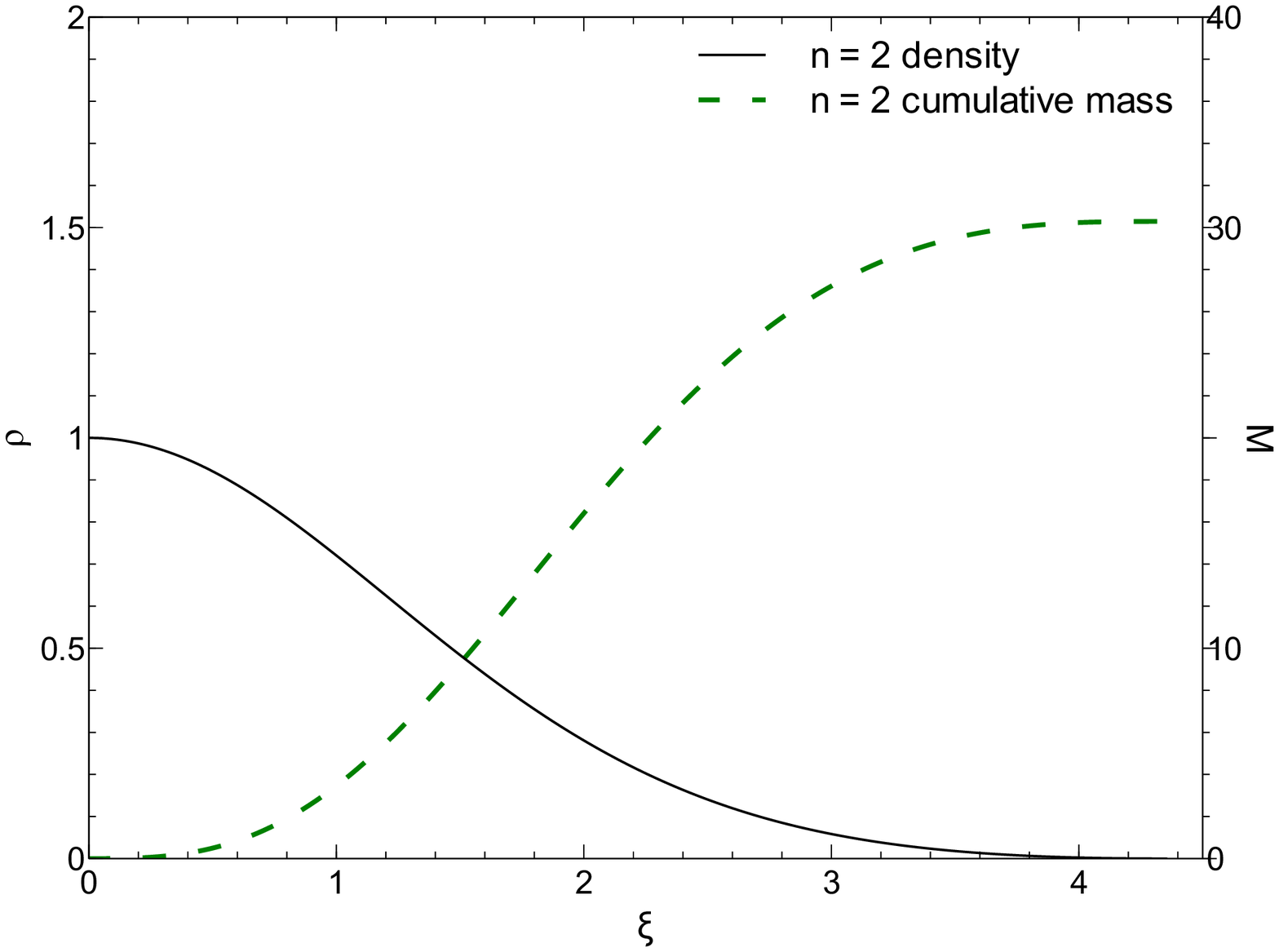}
\caption[]{Density profile and cumulative mass distribution
for the $n=2$ polytrope (in arbitrary units) which is the basis for the mapping between
$\rho$, $\Psi$ and $\Sigma$ in the polytropic cooling method. Note that the
mass distribution is weighted towards larger radius.}
\label{n2density.fig}
\end{center}
\end{figure}

It is worth noting some features of the approximation that will be useful in interpreting our results. Members of a polytropic family of given $n$ can all be mapped onto a single function representing the dependence of a scaled density-like variable on a scaled radius variable ($\xi$). The polytropic cooling  approximation first assumes that a given particle corresponds to a particular value of $\xi$ within a polytropic sphere and then, given the actual values of density and potential at that particle, determines the corresponding value that the column density would have in that case. This is repeated over a range of $\xi$ values and then a final `effective' column density is computed which weights the contributions from different $\xi$ in proportion to the amount of mass at different $\xi$ values within the polytropic sphere. Since for an $n=2$ polytrope, the majority of the mass is located at rather large radii  (see Figure \ref{n2density.fig}), this implies that the relationship between $\Psi$,  $\rho$ and $\Sigma$ is equivalent to assuming that the particle resides towards the outskirts of a structure. It  therefore implies a lower value of $\Sigma$ (for fixed $\Psi$ and $\rho$) than for a particle that was mapped onto the centre of a polytropic sphere. The polytropic cooling method 
applies this weighting over $\xi$ to all particles since, being entirely local, it has no knowledge of where any particular particle is located within a parent structure.  Therefore if, in fact, a particle is close to the core of its parent structure, the weighting of the polytropic cooling method will systematicallly underestimate the overlying column density. Conversely, for particles in the extreme periphery of a structure, the weighting of the polytropic  method over-estimates the column density at given $\Psi$ and $\rho$.

\begin{figure*}
\begin{center}
\begin{minipage}[]{16.4cm}
\includegraphics[width=5.4cm]{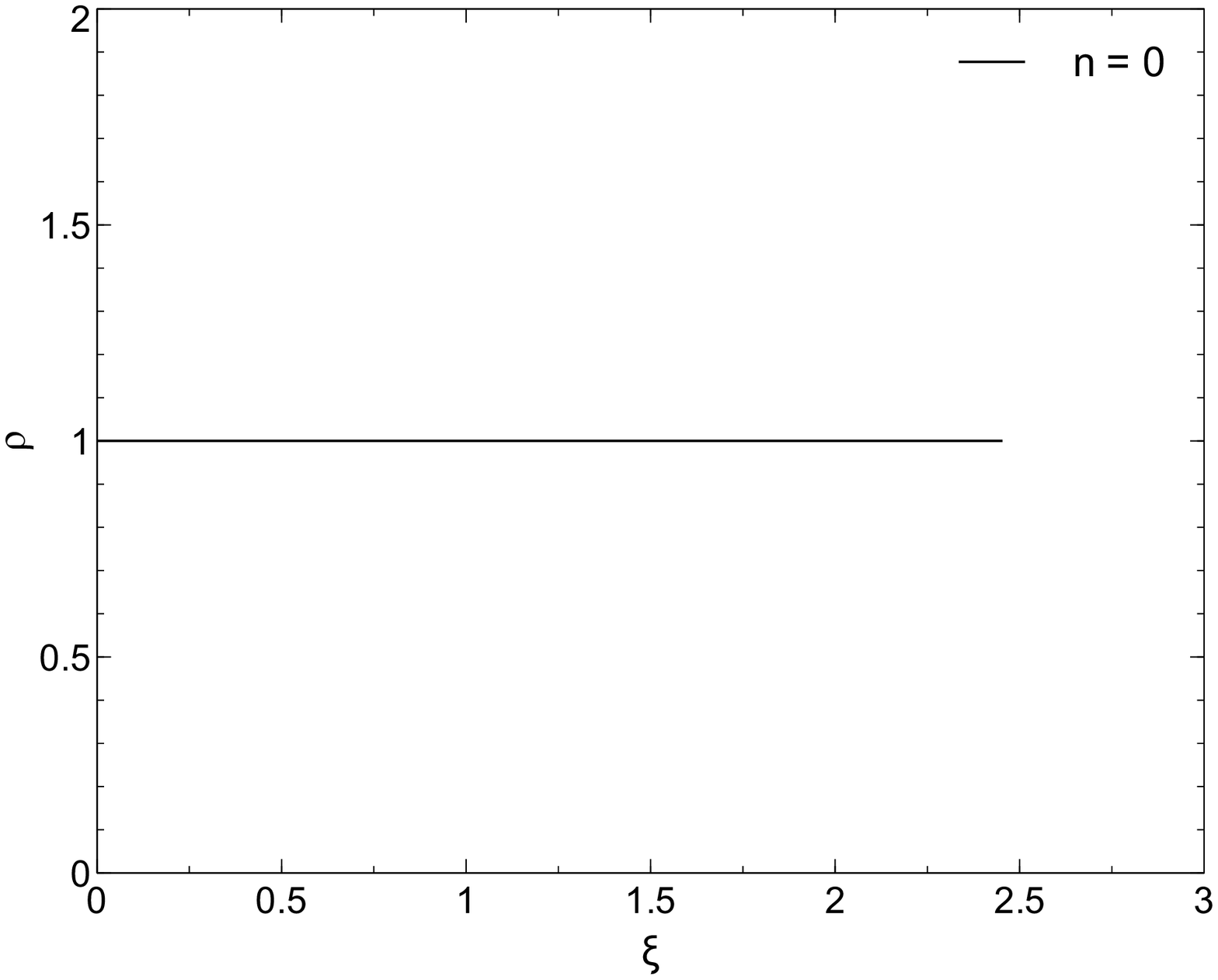}
\includegraphics[width=5.4cm]{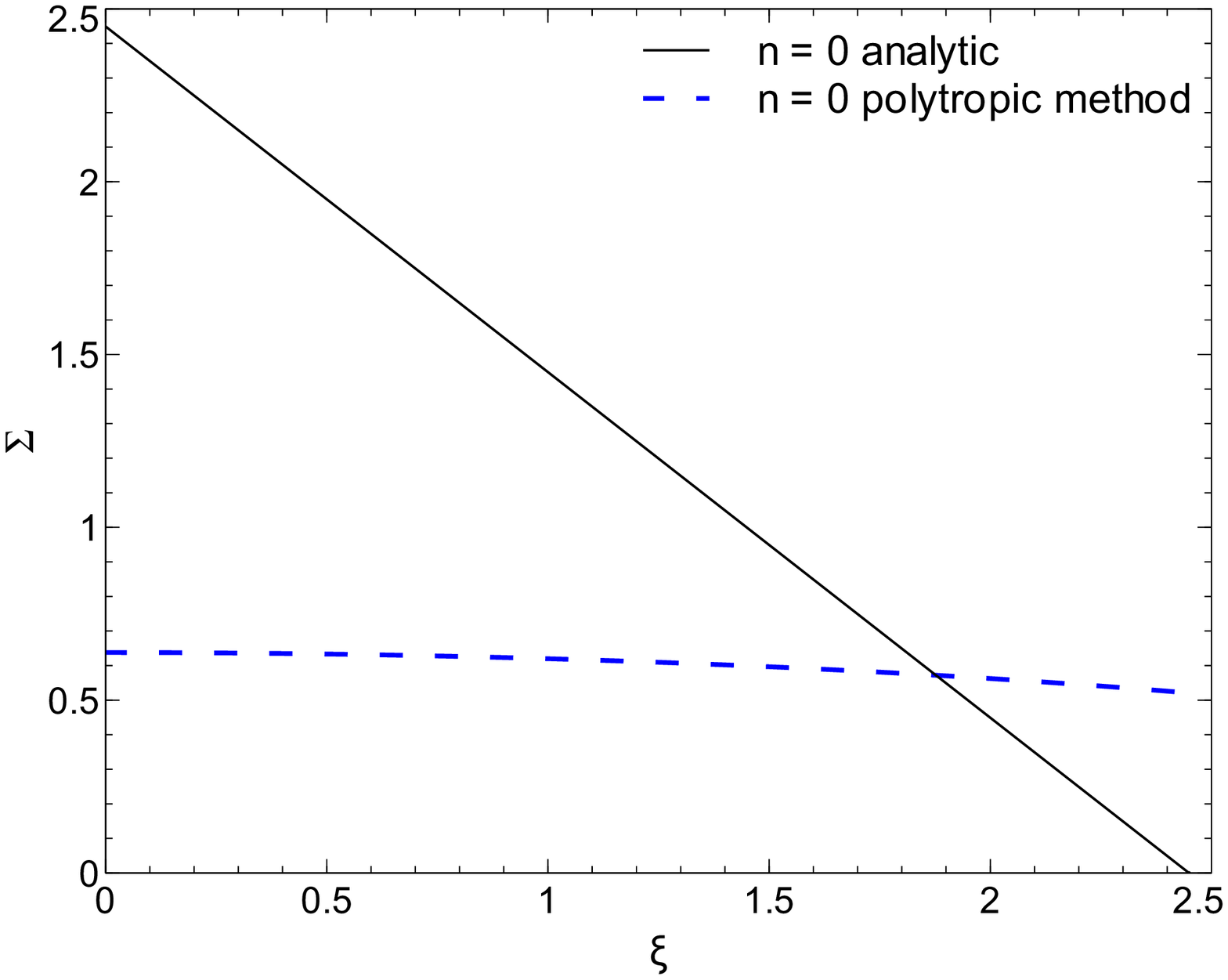}
\includegraphics[width=5.4cm]{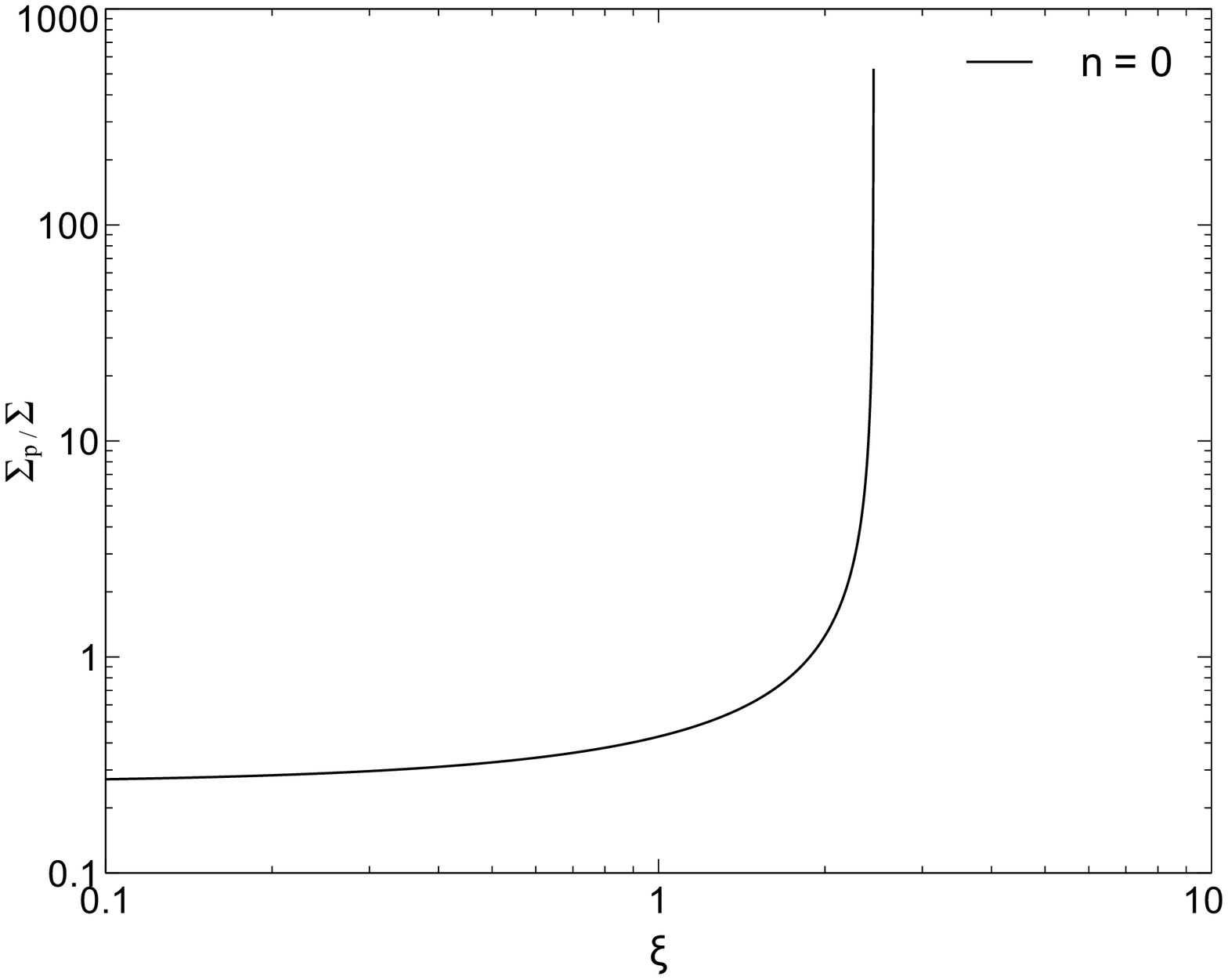}
\includegraphics[width=5.4cm]{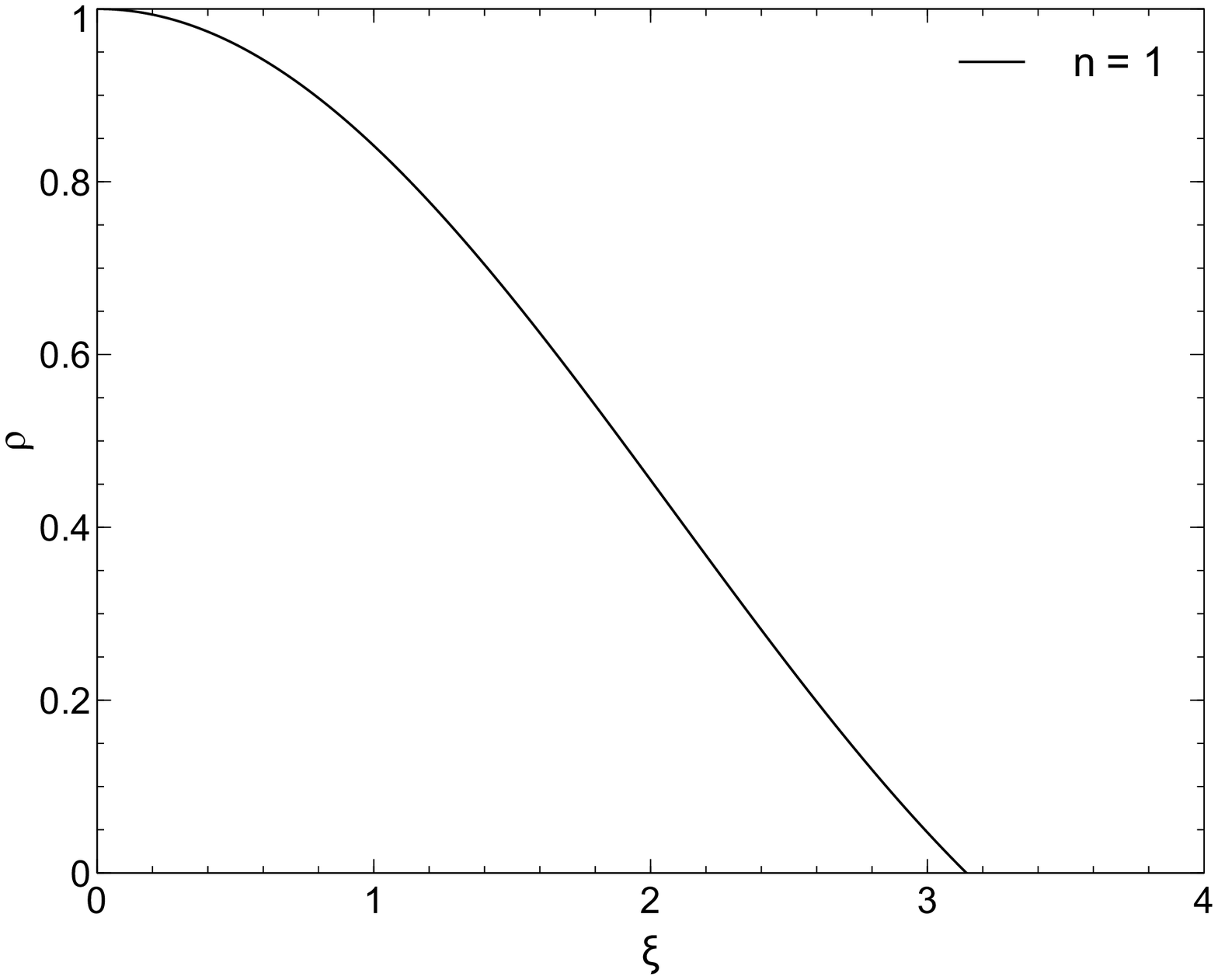}
\includegraphics[width=5.4cm]{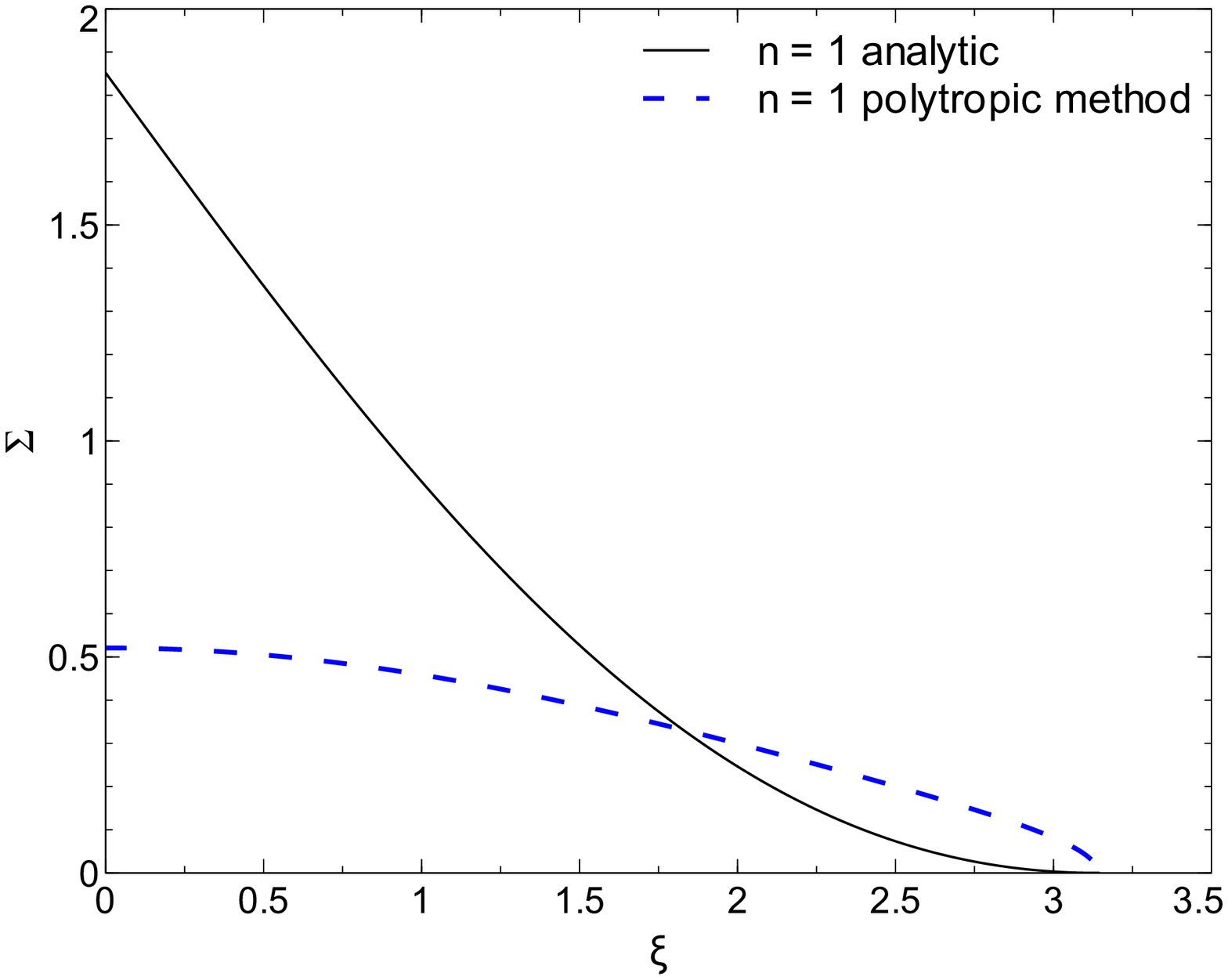}
\includegraphics[width=5.4cm]{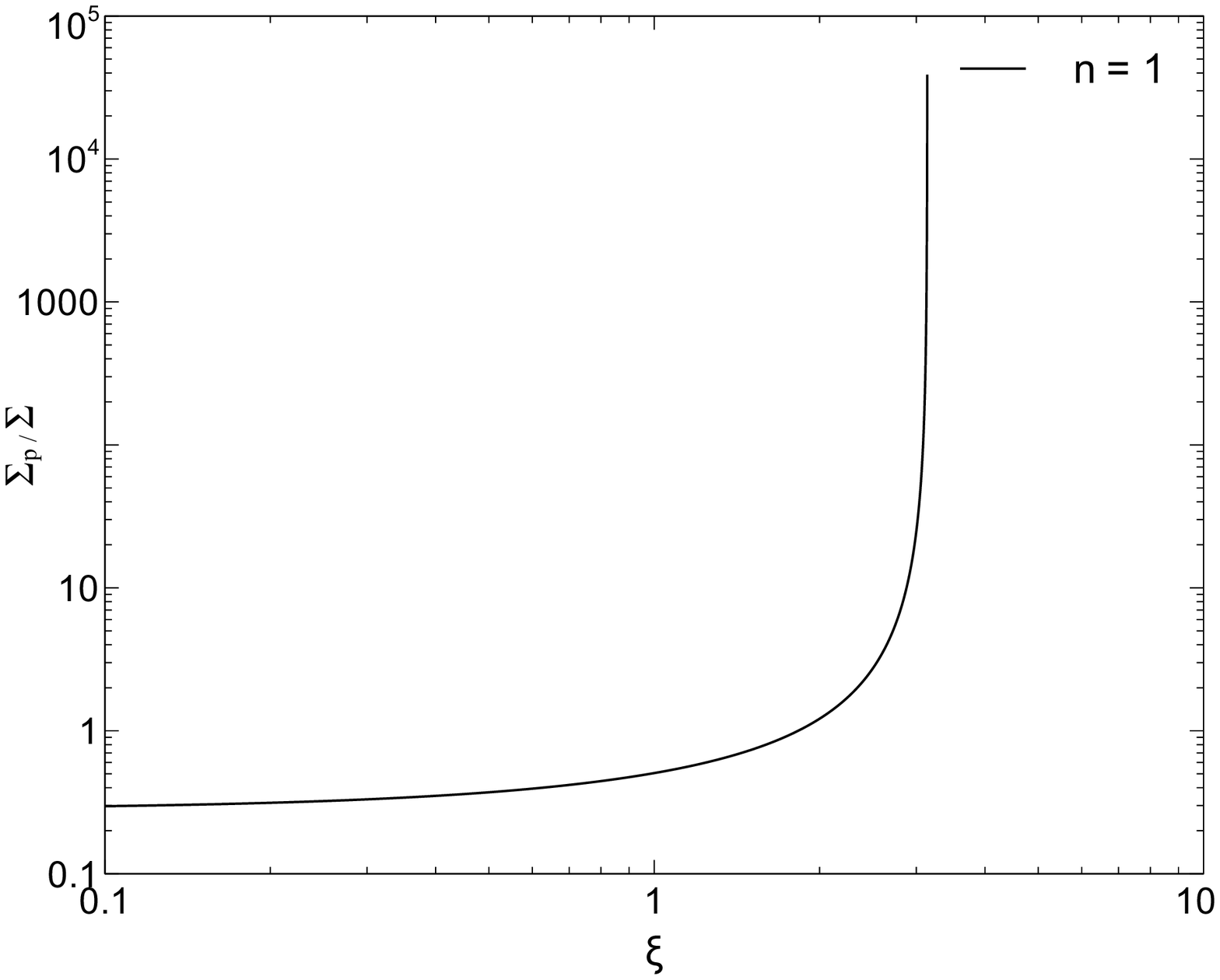}
\includegraphics[width=5.4cm]{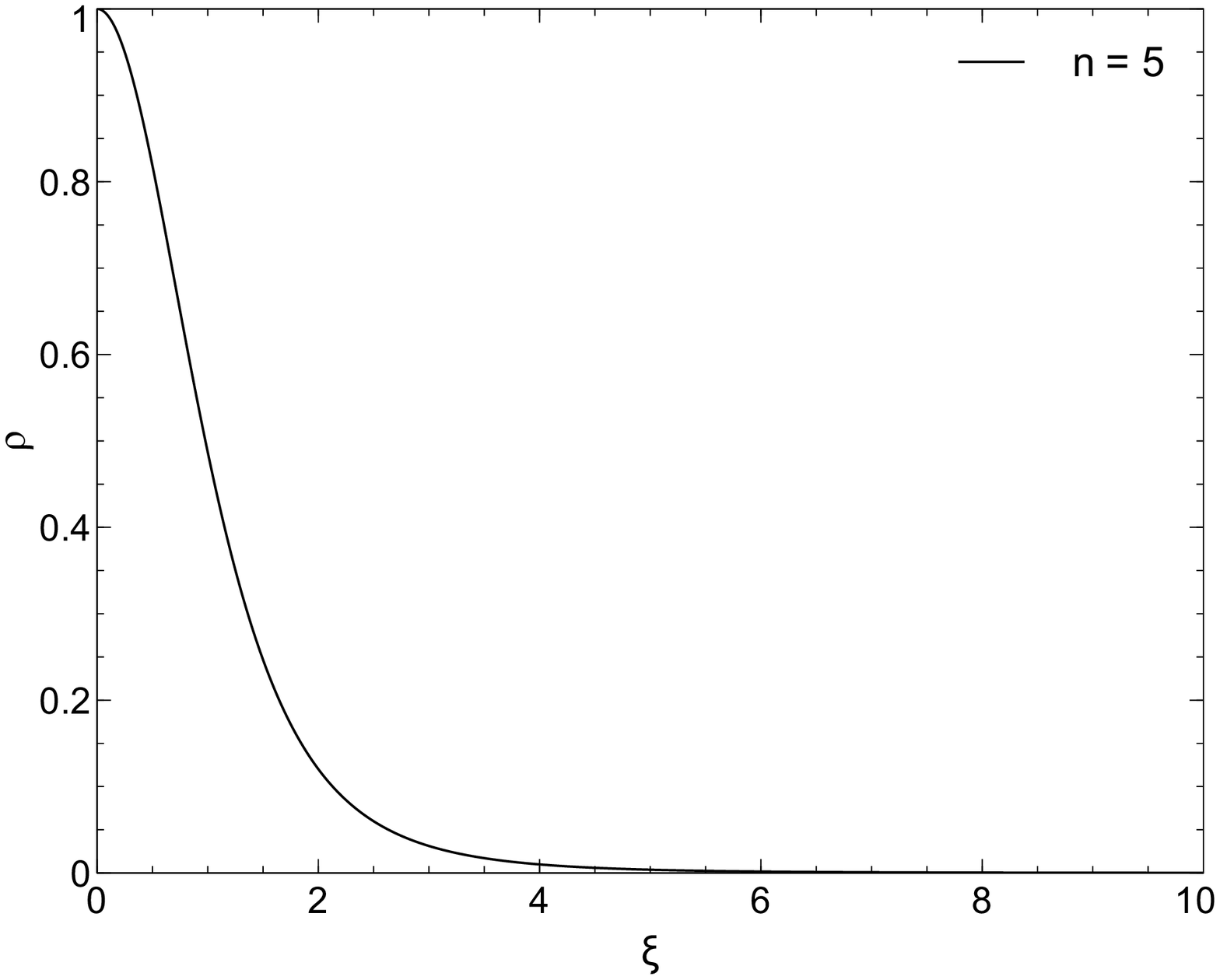}
\includegraphics[width=5.4cm]{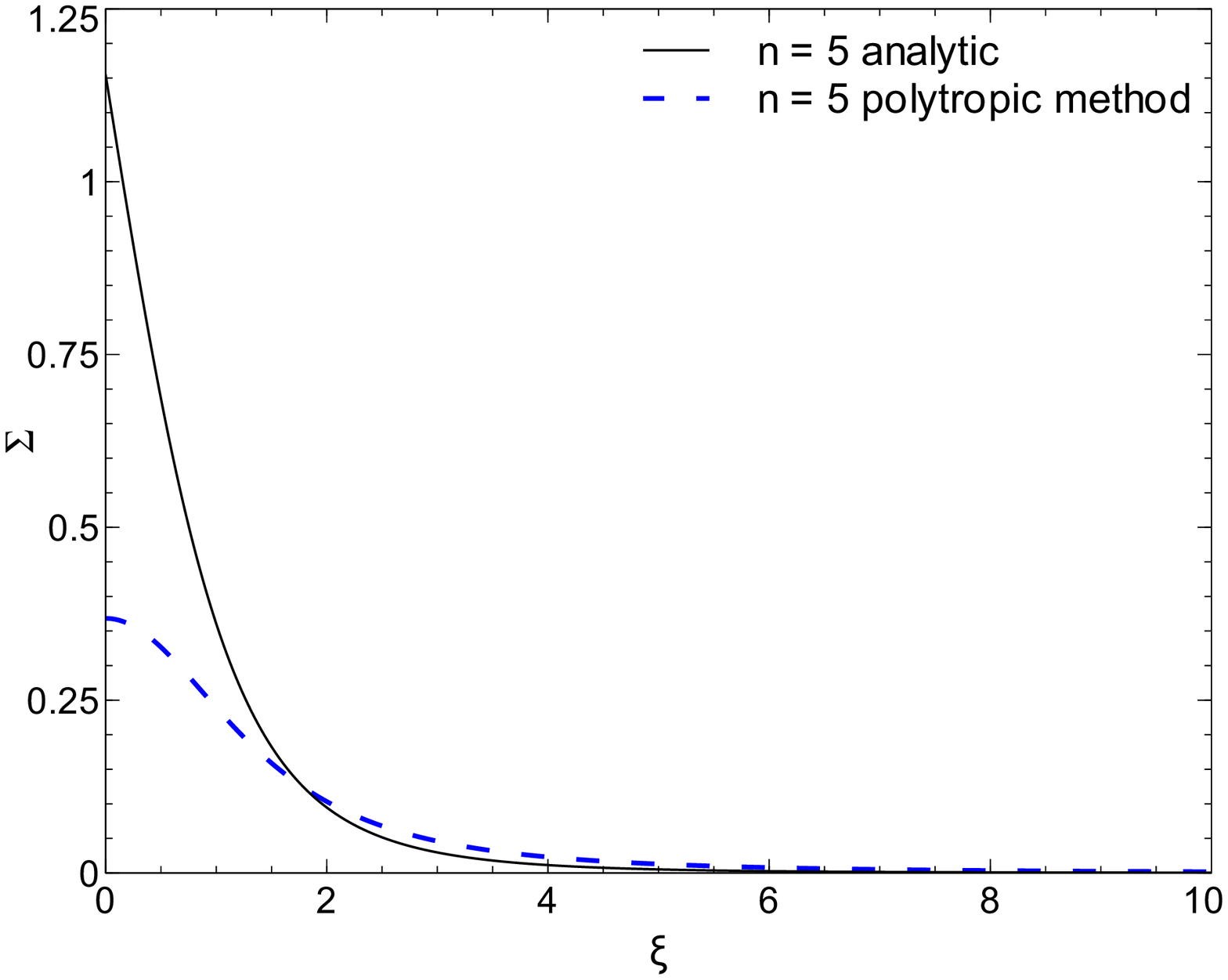}
\includegraphics[width=5.4cm]{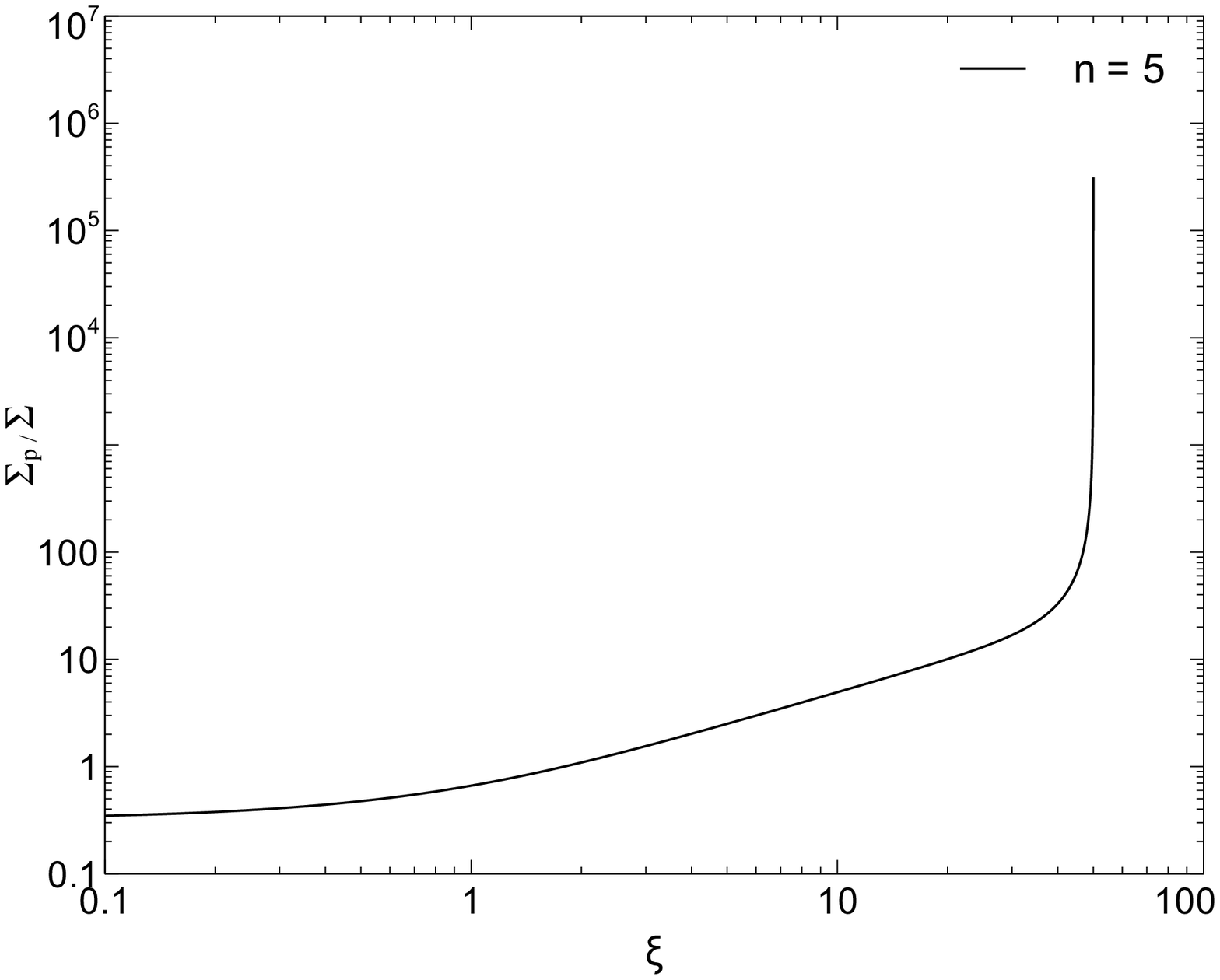}
\end{minipage}
\caption[]{Algorithm tests for spherical polytropes. Upper row:  $n=0$ (uniform density), middle row: $n=1$ polytrope, lower row: $n=5$ (nearly isothermal) polytrope. In each case, the left hand panel is the density profile, the middle panel compares the integrated column density to infinity from each point (solid line) with the corresponding quantity derived by the polytropic cooling  method (dashed line) and the right hand panel is the ratio of values for the column density on a logarithmic scale. Units are arbitrary.}
\label{polytropes_coldensity.fig}
\end{center}
\end{figure*}

This simple insight is sufficient for us to understand the method's systematic under-estimate of the column density near the centre of spherical structures and, conversely, its systematic over-estimate in the outer regions (Figure \ref{polytropes_coldensity.fig}). The method under-estimates the central column density by about a factor three for all polytropic spherical structures and the over-estimate at large radius is around a factor three for polytropes with $n$ of $2$ or greater and larger for smaller $n$ (less compressible gas).

\begin{figure}
\begin{center}
\includegraphics[width=8cm]{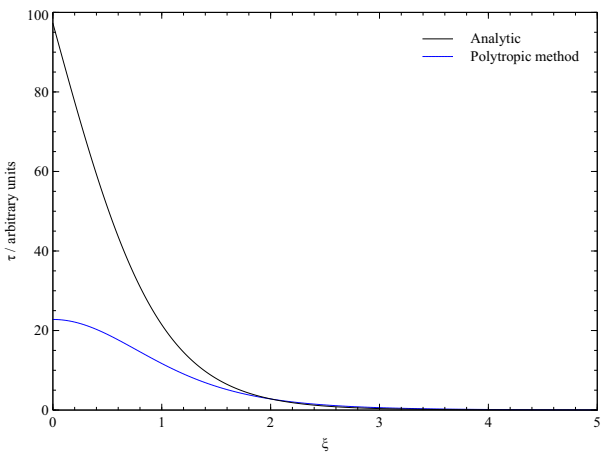}
\caption[]{Comparison of the optical depth (in arbitrary units) due to ice and the polytropic  estimate of the same quantity (dashed line) in the case of the $n=4$ polytrope. Note that the normalisation for the optical depth is arbitrary.}
\label{tau_ice.fig}
\end{center}
\end{figure}

 In the case of constant opacity (\textit{e.g.} that due to electron scattering), these results for the column density apply directly to the derived  optical depth since the two quantities are simply proportional. 
In Figure \ref{tau_ice.fig} we present the optical depth as a function of radius for the
case of opacity given by ice grains for which $\kappa \propto T^2$.
Note that normalisation of optical depth is arbitrary.  This illustrates that within the region containing the bulk of the mass, the polytropic method 
underestimates the optical depth.

\subsection{The application of the polytropic method to disc-like geometries}
\label{disc.sec}

 We  test the polytropic  method in planar geometry, being mindful of the fact
 that this  applies to a number of situations that occur in hydrodynamical simulations such as shock compressed sheets and rotationally supported discs.\footnote{Note that although \citet{stamatellos_whitworth} demonstrated that
the method satisfies the Hubeny test \citet{hubney} for the variation of
temperature as a function of optical depth within a disc, this does not
demonstrate that the method has calculated the optical depth correctly, which is the focus of our investigation here.}  To this
 end we use an approximately Keplerian  disc model in which both the column
density ($\Sigma$) and vertically averaged temperature vary as $R^{-3}$.
Such a structure has a radially constant profile of Toomre $Q$ parameter
\citep{toomre} as would be expected in the case of discs that are in
a state of self-regulated marginal stability against self-gravity
\citep{gammie,rice}. In fact, this particular profile corresponds to
the case of a disc with constant accretion rate and radiative cooling with
opacity provided by ice grains \citep{clarke-09}, an opacity regime that
is applicable to the outer regions of discs around young stars and where
the cooling rate is important in determining whether such discs fragment.

  Although (as emphasised by Stamatellos et al 2007), their cooling
prescription is not only applicable to situations that are in hydrostatic
equilibrium, we here test the method on a density and temperature field
that most closely resembles the situations that arise in star formation
simulations in which hydrostatic equilibrium normal
to the disc plane is attained on a short
(dynamical) timescale (we assume that the small influence of pressure
gradients and of the disc's self-gravity in the disc plane is
balanced by a minor adjustment of the disc's rotation curve). In order
to obtain an analytic approximation to the disc vertical structure as
a function of radius, we approximate the vertical gravitational
acceleration at height $z$ in the disc as
\begin{equation}
g_z = -\Omega^2 z - 4 \pi G \Sigma ^\prime
\label{gz}
\end{equation}
where the first term is the vertical component of the gravitational
acceleration provided by the central star ($\Omega$ is the local
Keplerian angular velocity)  and $\Sigma ^\prime$ is the column
density {\it between the point at height z and the disc mid-plane}
(which is not to be confused with the column density to infinity that
has been discussed hitherto). This second term is the gravitational
acceleration that the disc would exert in the limit that it was
an infinitely extended stratified slab. In reality there are further
small corrections to the self-gravity  due to radial gradients in the disc
(see {\it e.g.} the appendix of \citealp{bertin_lodato}). We have verified
{\it a posteriori} (by subtracting the density distribution of a
radially constant disc from the computed structure and then calculating
$g_z$ by direct summation) that this
further correction would account for at most
$\sim 3 \%$  of the total $g_z$ value. We emphasise that this approximate
$g_z$ value is only used in order to obtain a structure that is
in approximate vertical hydrostatic equilibrium and that the full disc
(and star) potential is used when computing the polytropic cooling.

  The motivation for adopting this approximate form of $g_z$ is that
it allows us to compute an {\it analytic disc structure as a function of
cylindrical radius ($R$) and z}, since such a disc can be modeled by
an $n=2$ polytrope with $P=K\rho^2$ where $K$ is a global constant.{\footnote{Note that although this equation of state yields the 
surface density profile of a steady state  marginally self-gravitating disc 
with opacity dominated by ice cooling, it does not do so uniquely, since the
actual pressure-density relationship in such a disc depends on issues such
as the vertical profile of viscous dissipation in the disc. It nevertheless
offers a simple analytic desription of a disc with an astrophysically
relevant surface density profile.}}
In this
case, differentiation with respect to $z$ of the hydrostatic equilibrium
equation normal to the mid-plane yields the modified simple harmonic
equation:
\begin{equation}
{{\partial^2 \rho}\over{\partial z^2}} = -{{2 \pi G}\over{K}} \bigl(\rho + {{\Omega^2}\over{4 \pi G}} \biggr)
\label{rhode}
\end{equation}
whose solution (subject to $\rho = \rho_0$ and $\rho^\prime = 0$ at
$z=0$) is
\begin{equation}
\rho=\rho_0\biggl(\bigl(1+ f(Q)\bigr){\rm cos} {{z}\over{H_0}}
-f(Q)\biggr)
\label{rhosolve}
\end{equation}
where
\begin{equation}
f(Q) = {{\Omega^2}\over{4\pi G \rho_0}}
\label{fdef}
\end{equation}
and
\begin{equation}
H_0 = \biggl({{K}\over{2 \pi G}}\biggr)^{1/2}
\label{h0def}
\end{equation}

 The constant $f$ is of order $Q$ when $Q \gg 1$ (the non-self gravitating
case) while $f \sim Q^2$ in the strongly self-gravitating case
($Q \ll 1$). In the marginally self-gravitating (self-regulated) case
that we consider here, $f$ is of order unity.

The density profile given above allows us to derive the scale height
of the disc (defined here as being the height at which the density
falls to zero):
\begin{equation}
H = H_0 {\rm cos}^{-1} \biggl({{f(Q)}\over{1+f(Q)}}\biggr)
\label{hdef}
\end{equation}

\begin{figure}
\centering
\includegraphics[width=8.2cm]{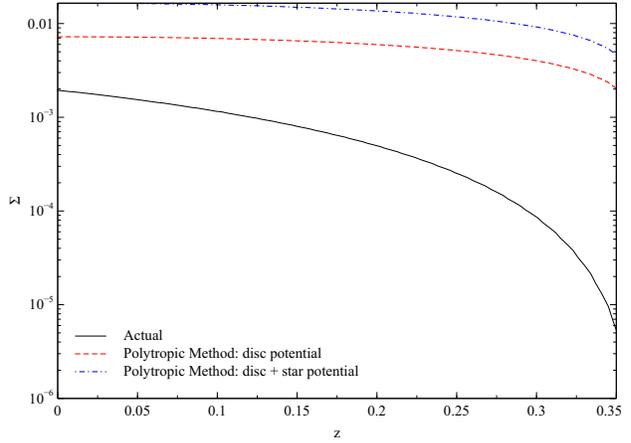}
\caption[]{The column density in the $z$ direction out to infinity in disc geometry, compared to the polytropic estimate.} 
\label{disc_sigma.fig}
\end{figure}

We set up such a disc over the radial range $R=1$ to $R=10$
(in arbitrary units), and adopt parameters $f=1$ and $H_0 = 0.35$.  
For this model  the total disc mass (which, given the surface density profile,
is concentrated at small radii) is $\sim 20 \%$ of the central star mass.
At given $R$ and $z$, we compute the gravitational potential
by adding the contributions due to the star and the rest of the disc. The
latter is calculated by using elliptic integrals to compute the
contribution to the potential from each ring of
material corresponding to each pair of $R$ and $z$ values in the grid.
The local density and potential is then used to obtain the polytropic
cooling method's estimator of the column density which is compared with the
`true' column density integrated out to infinity in the $z$ direction. Figure \ref{disc_sigma.fig} compares these column density values as a function of $z$ for a
cylindrical slice at $R=5$, at which point the disc aspect ratio
$H/R \sim 0.07$. The upper curve in the Figure represents the case that
the full (star plus disc) potential is used in the estimation of
$\Sigma$, whereas the intermediate curve uses just the disc potential
(an approach suggested by Stamatellos et al 2007 though not always implemented
in disc studies). As expected, the estimated column density is larger
when the stellar potential is included by around a factor $2\sim 3$, consistent
with the square
root dependence of the estimated column density on the potential and the
fact that the disc mass is concentrated towards the inner part of the disc. 
{\it However, even in the case that only
the disc potential is included, the estimated column density exceeds the
true column density by a factor 5 or more at all heights.} The consistent
sign of the discrepancy (in the sense that  the estimated column is always larger)
is simply because in a disc geometry there is always a low column density
route to infinity (along the symmetry axis) which is not accounted for
by the estimated value.

\subsection{Summary}
We find that in the case of spherical structures, the polytropic 
method under-estimates the column density (and hence optical depth
in the case of constant opacity discs) by a factor of $3\sim4$ in the
inner regions and over-estimates the column density at large radius.
This can be understood in terms of the weighting function that is used
in order to map the density and potential onto the column density which
is weighted towards the relationship between these quantities that
applies at intermediate radii within an $n=2$ polytrope. 

On the other hand, the method {\it systematically} over-estimates
the (vertical) column density in a disc structure, even in the case that
one uses only the potential from the disc (rather than the
disc plus star) in estimating the column density.
For the disc structure we have modeled
(for which $H/R \sim 0.07$ at the radius considered
and where $Q_T \sim 1$ throughout), this over-estimate is around a factor $5$
or more.
This over-estimate  can be understood as a geometric effect
in that the estimate based on local density and potential does
not take account of the direction of low $\Sigma$ normal to the disc plane.

\section{Implications of results for the estimate of cooling rates}
\label{cooling.sec}

 The polytropic  method uses the column density and optical depth
estimates discussed above to compute the cooling rate per unit mass
according to
\begin{equation}
\label{cooling_approx.equ}
\dot U = \frac{\sigma T^4 - \sigma T_0^4}{\bigl( \tau + \tilde \tau^{-1}\bigr)\Sigma}
\end{equation}
Here, $\tau$ is the path averaged optical depth (see Appendix \ref{taubar.sec})
whereas $\tilde \tau$ is the product of local opacity and estimated column
density. 
These two quantities differ by a factor that is usually of order unity, except
in regions where the opacity is strongly dependent on temperature (see Figure 6 of
\citealp{stamatellos}). Note that in both cases the opacity is given by
parametrisations of the Rosseland mean opacity
 
 There are several approximations involved in Equation \ref{cooling_approx.equ}. First of all, it is assumed that the cooling rate per unit mass at a given point is simply an estimator of the local radiative flux divided by the estimated column of material above this point. Secondly, it takes a form for the local radiative flux designed to have particular  functional forms in the limits of small and large optical depths. At low optical depths this is simply the product of the optical depth and the black body flux. In the limit of large optical depth, this expression is motivated by the radiative diffusion approximation

\begin{equation}
\label{flux_diff.equ}
F = -{{4}\over{3 \kappa \rho}} \nabla (\sigma T^4)
\end{equation}
and is likely to be of similar order to Equation \ref{flux_diff.equ}, provided that all quantities vary over a scale length $r$ (so that $\frac{T^4}{\tau} \sim \frac{dT^4}{d\tau}$). Finally, the second term in the numerator of Equation \ref{cooling_approx.equ} ensures that the cooling rate goes to zero at temperature $T_0$ and is designed to mimic the effect of an external radiation field at this temperature. For the purposes of the test we set $T_0 = 0$.

  However, it is important to recognise that Equation \ref{cooling_approx.equ} is itself an approximation. In the optically
thick limit we can also evaluate the `true' cooling rate per unit mass, using

\begin{equation}
\label{cooling_diff.equ}\dot U_{thick} = {{1}\over{\rho}} \nabla \cdot F
\end{equation}
where $F$ is evaluated in the diffusion limit (\textit{i.e.} Equation \ref{flux_diff.equ}).
\subsection{Cooling rates in spherical geometry}

\begin{figure*}
\begin{center}
\begin{minipage}[]{14cm}
\includegraphics[height=7cm,angle=270]{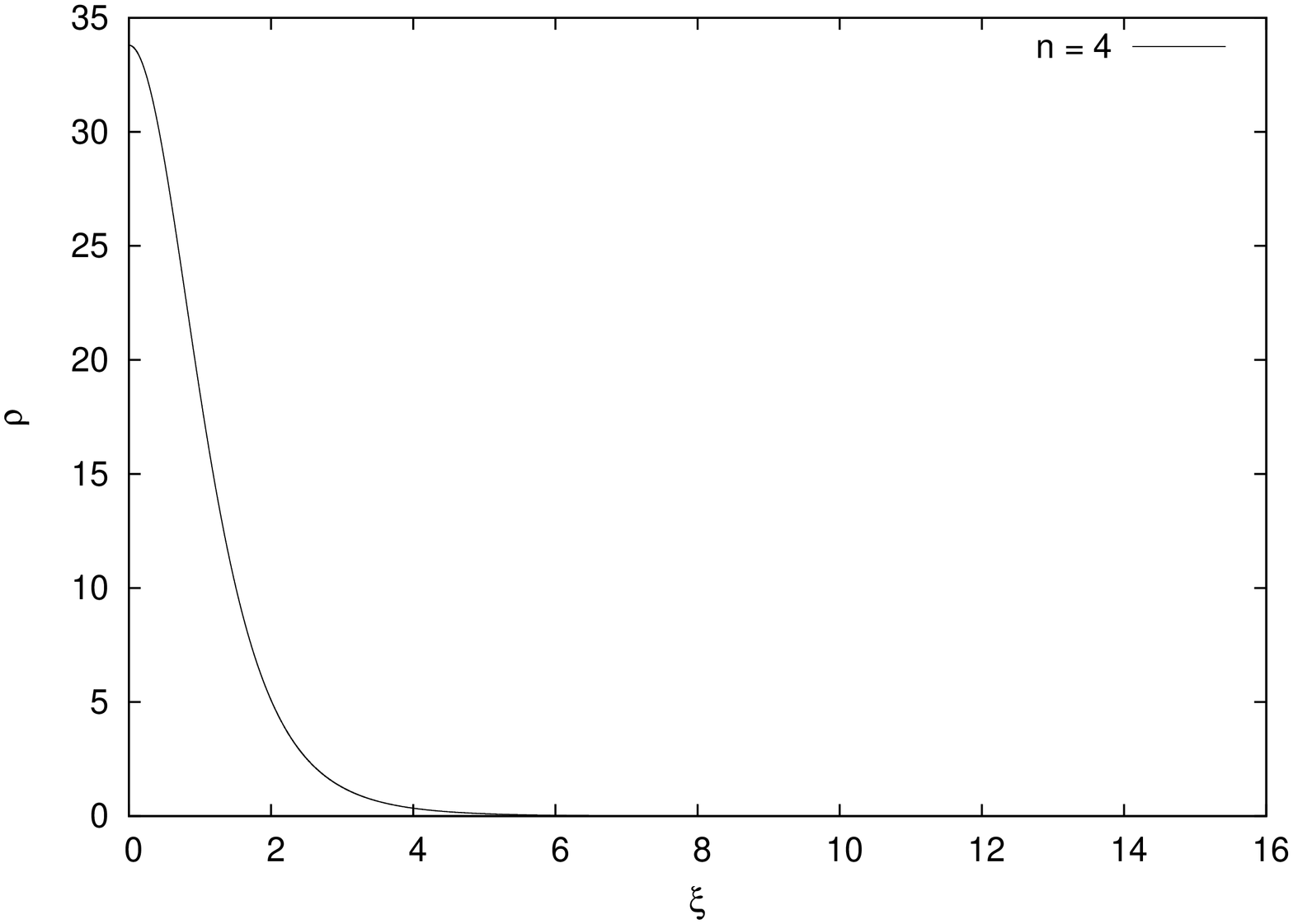}
\includegraphics[height=7cm,angle=270]{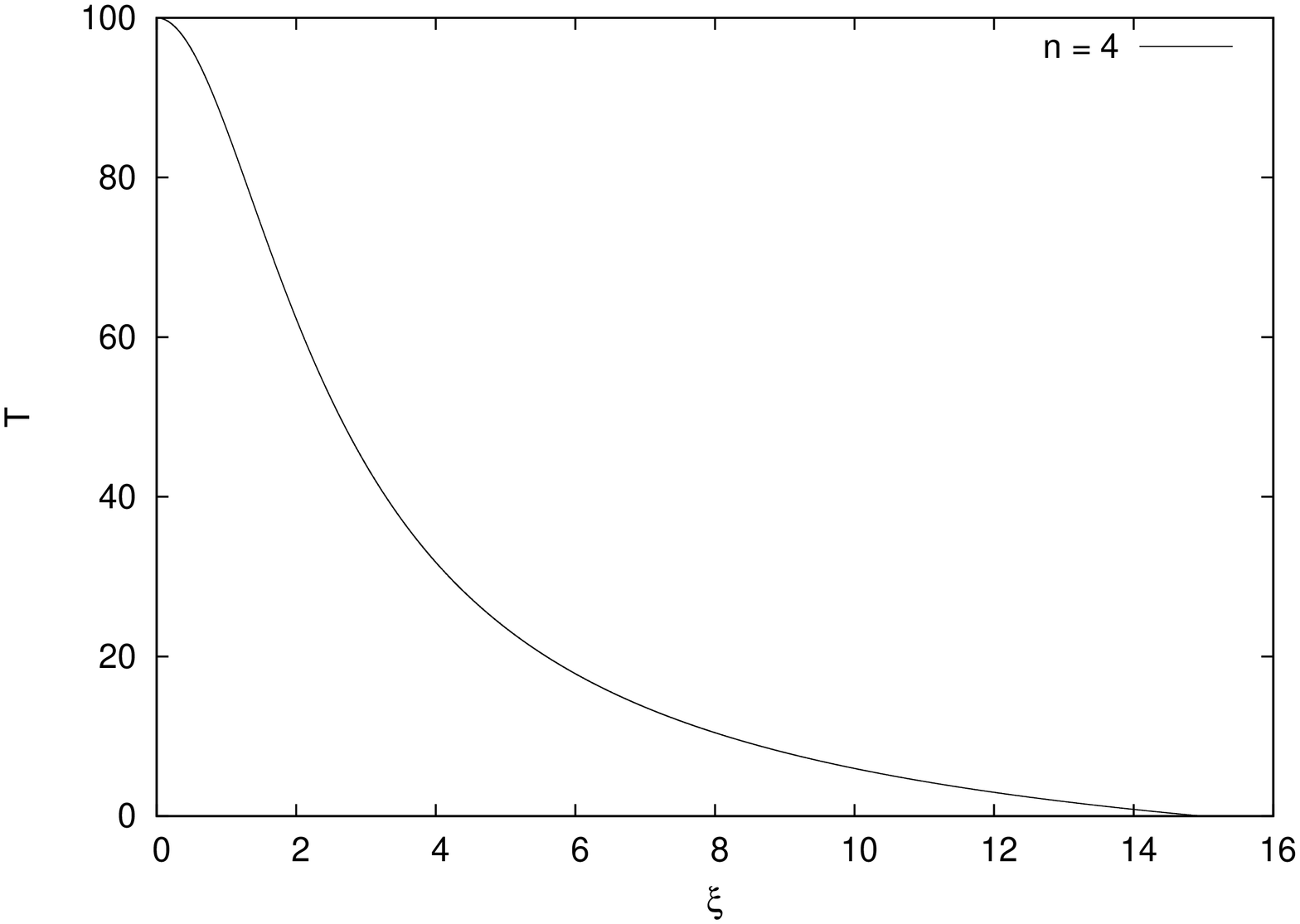}
\end{minipage}
\caption[]{Density (left panel) and temperature (right panel) profiles, in arbitrary units, of the hydrostatic $n=4$ polytrope used for the cooling rate comparisons.}
\label{n4.fig}
\end{center}
\end{figure*}

\begin{figure*}
\centering
\begin{minipage}{170mm}
\subfigure[]{
\includegraphics[width=8.5cm]{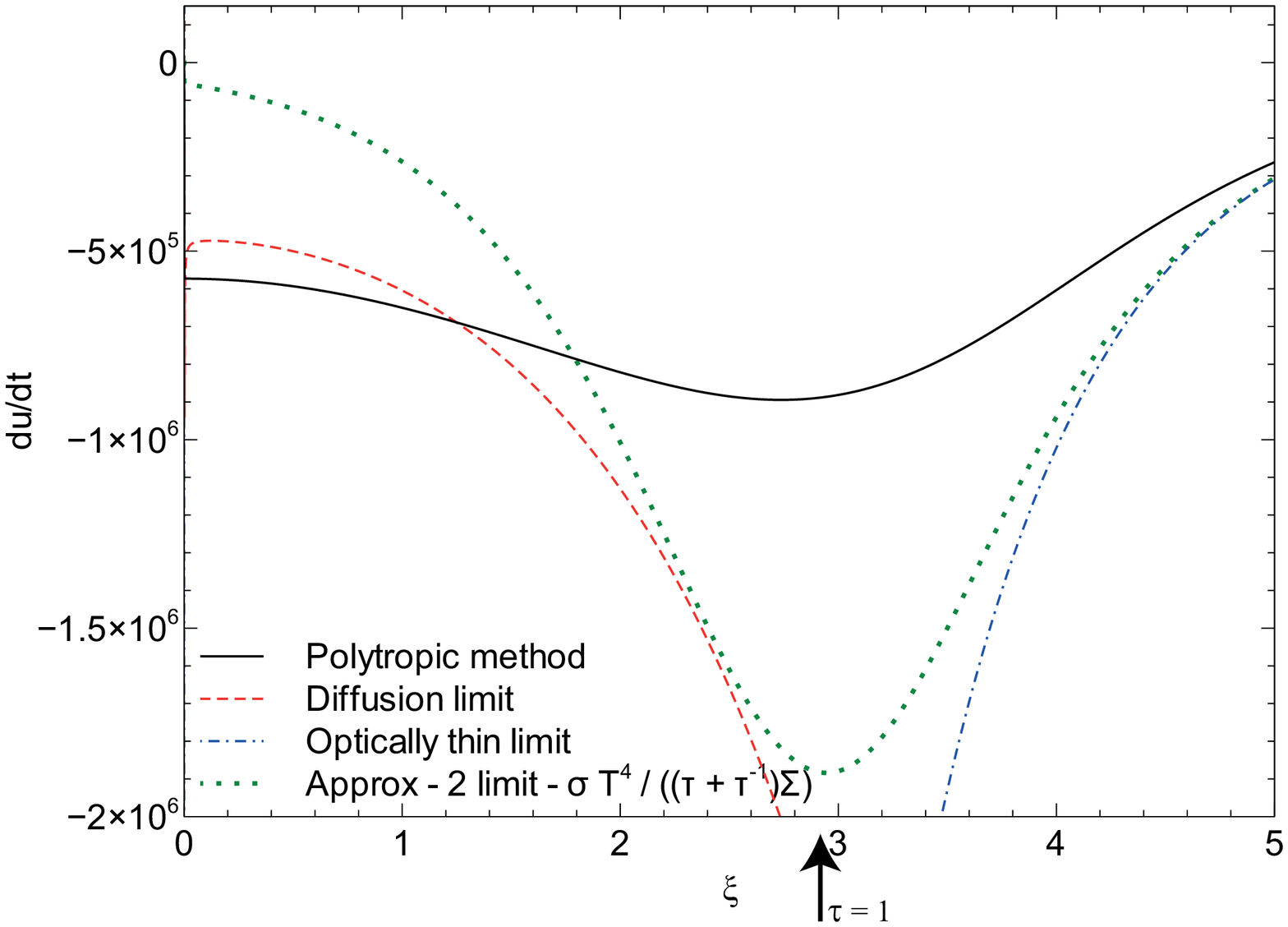}
\label{polytrope_cooling.fig:const}
}
\subfigure[]{
\includegraphics[width=8.5cm]{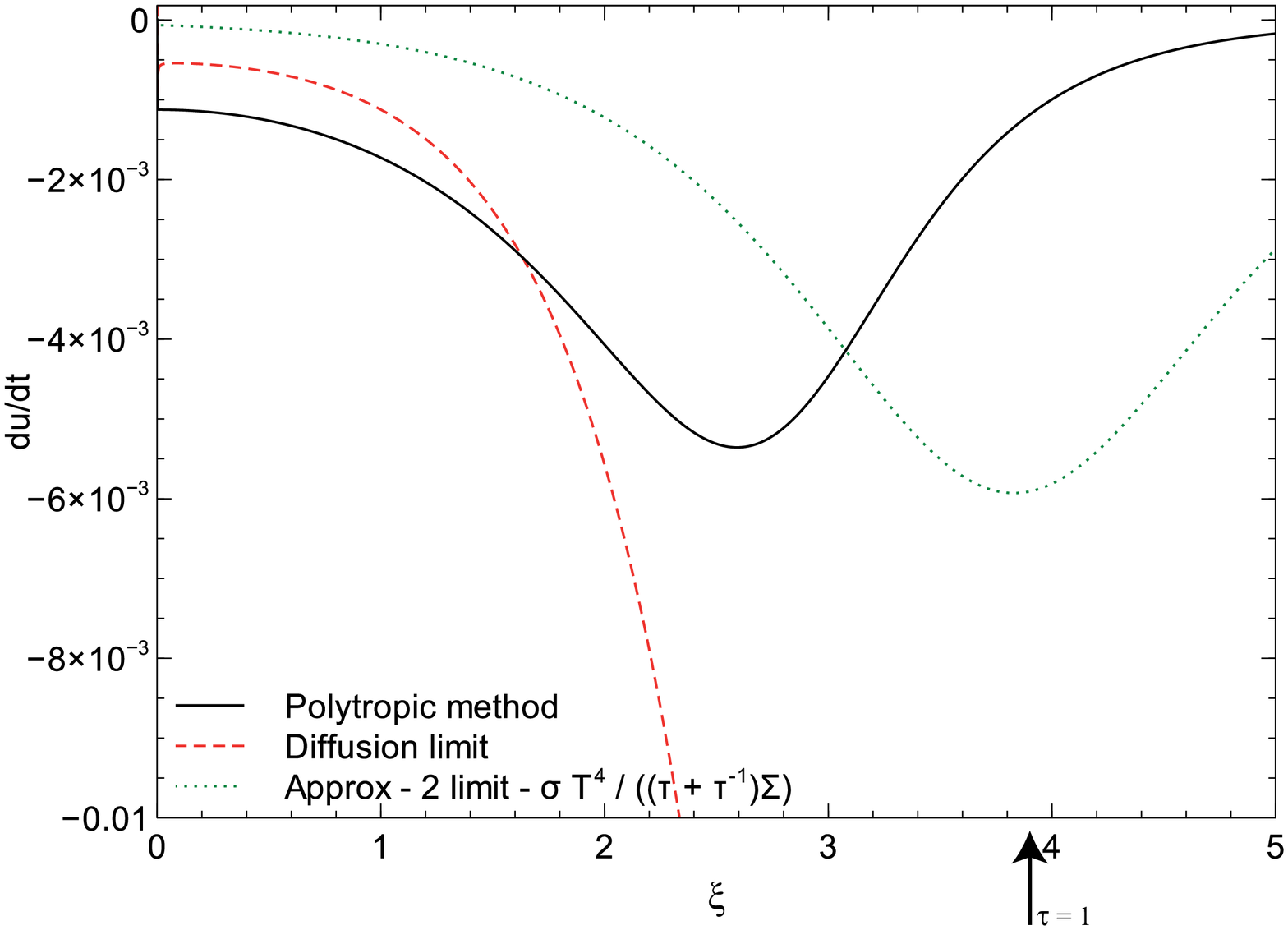}
\label{polytrope_cooling.fig:ice}
}
\caption[]{Comparison of estimates of the cooling rate per unit mass (in arbitrary units) for the case of a) constant opacity and b) ice grain opacity for
$n=4$ polytropic spheres with (true) central optical depths of $42$ and $100$, respectively. The black line represents the polytropic estimate to Equation \ref{cooling_approx.equ} while the green dotted
line uses the same equation but with the true values of the optical depth and column density. The blue dash-dot line is the limiting
form of both expressions at low optical depth where they agree by construction. The red dashed line is the `true' cooling rate computed using the radiative diffusion approximation (Equations \ref{flux_diff.equ} and \ref{cooling_diff.equ}).
The arrows denote the point of (true) optical depth equal to unity.}
\end{minipage}
\label{polytrope_cooling.fig}
\end{figure*}

We first examine how the cooling values  predicted by Equation \ref{cooling_approx.equ} using column density and optical depths estimated by the polytropic
method compares with the same quantity evaluated using the actual optical
depths and column densities. In each figure, the left hand panel refers to the case of constant opacity and the right hand panel to the case $\kappa \propto T^2$.
For the purpose of this comparison, we evaluate the temperature profile (Figure \ref{n4.fig})
by assuming that the structure is in hydrostatic equilibrium. The (true) optical depth to the centre is $42$ for the constant opacity case and 100 for the case of ice and the point at which the (true) optical depth is  unity is marked by the arrow in Figure \ref{polytrope_cooling.fig}.     

Figure \ref{polytrope_cooling.fig} shows that the polytropic cooling  estimate of Equation \ref{cooling_approx.equ} (black line) has the same
general shape as the same quantity evaluated from the actual column density (green dotted line), in that in both
cases the cooling rate per unit mass is highest close to an optical depth of unity and that both
converge to the same value in the limit of low optical depth. This is to be expected since as $\tau$ tends to
zero, the cooling rate per unit mass tends to a value that is independent of column density and depends only
on temperature (the optically thin limit of Equation \ref{cooling_approx.equ} is shown by the blue dash-dot line in
Figure \ref{polytrope_cooling.fig}). However, in the optically thick portion of the disc, the two cooling rates are in general quite
different, with the polytropic  estimate over-estimating the cooling at small radii and under-estimating
it at large radii. This can be simply understood in that the cooling rate per unit mass
in the optically thick limit scales as $1/\Sigma^2$ and hence this result reflects the respective
under-estimate and over-estimate of $\Sigma$ at small and large radius. 

  So far, we have just examined how the accuracy of estimating the column 
density affects the cooling rate as specified by Equation \ref{cooling_approx.equ}. A more important comparison is however the comparison of the polytropic
method's estimated cooling rate with the `true' (radiative diffusion)
method in the limit of high optical depth (Equations \ref{flux_diff.equ} and \ref{cooling_diff.equ}),
shown as the dashed red line in Figure \ref{polytrope_cooling.fig}.
Interestingly, this agrees rather well with the polytropic method in
the optically thick region (the black line) in both opacity regimes.
In both cases the total (mass integrated) cooling rates agree to within
a few tens of per cent.
Such a discrepancy
is almost certainly no larger than additional errors due to poorly determined
opacities, for example. We have also explored a variety of different
polytropic spherical structures and different opacity laws and find that the
polytropic cooling method's fair (i.e. order unity)
agreement with the results of radiative
diffusion is reproduced for a wide range of input parameters. This explains why the method performed well in reproducing the results of the spherical collapse
calculation of \citet{masunaga}.

When the cooling rates are integrated over mass through the optically thick part of the structure, one obtains the result that the polytropic method modestly under-estimates the total cooling
rate by around $20 \%$ (compared to the radiative diffusion limit) in both cases. The switch in the sign of the discrepancy at intermediate radius means that
the global values agree rather well despite the discrepancies in local rates.

  It is perhaps surprising that the polytropic cooling method reproduces
the `true' (radiative diffusion) cooling rate considerably better than it
reproduces the quantity that it ostensibly computes (Equation
\ref{cooling_approx.equ}). This is because Equation \ref {cooling_approx.equ}
is only a reasonable approximation of Equation \ref{flux_diff.equ} in the 
case that one can replace differentiation with respect to $\tau$ by
simple division by $\tau$. Comparison between the green and red lines
in Figure \ref{polytrope_cooling.fig}  show that this is a poor approximation
at small radius, {\it i.e}. the temperature falls over an optical depth that is a 
small fraction of the total optical depth at small radius and hence simple
division by $\tau$ under-estimates the magnitude of the cooling. This
is almost exactly compensated by the fact that the polytropic cooling method
under-estimates the optical depth in the central regions (see \S\ref {coldensity.sec}). Although it may be regarded as 
unfortunate that the reason for this good agreement is a near cancellation of
discrepancies of comparable magnitude, we are encouraged that,
as described above, the reasonable agreement appears to hold  - {\it
in spherical geometry} for
a wide range of input parameters.  

\subsection{Cooling rates in disc geometry}

\begin{figure}
\centering
\includegraphics[width=8.2cm]{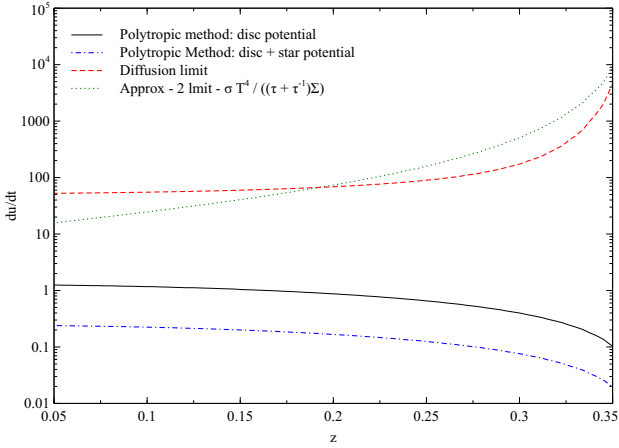}
\caption[]{The cooling rate for the disc system evaluated by the polytropic cooling method (both including and excluding the stellar potential) compared with the diffusion limit as well as Equation \ref{cooling_approx.equ} using analytic optical depths at radius $R=15$.} 
\label{disc_cool.fig}
\end{figure}

  The situation in disc geometries is very different. As above, we start
by comparing the polytropic cooling estimate of Equation \ref{cooling_approx.equ} 
with the same quantity evaluated using the true column density at each point.
For consistency with the cooling regime encountered in the outer regions
of self-gravitating discs, we employ the opacity appropriate to ice cooling
($\kappa \propto T^2$).
We here focus on the optically thick regions of the disc, in order to
allow comparison with Equation \ref{flux_diff.equ} and so neglect the additive term
$\propto \tilde \tau^{-1}$ in the denominator of Equation \ref{cooling_approx.equ}. Figure \ref{disc_cool.fig} plots the specific
cooling rate estimates as a function of $z$ for the slice at $R=5$ 
in the model disc described in \S\ref{disc.sec}. The dashed line represents the cooling
rate according to the radiative diffusion approximation ({\it i.e.} Equations
\ref{flux_diff.equ} and \ref{cooling_diff.equ}) while the upper solid line employs Equation \ref{cooling_approx.equ} where 
 the column density is the `real' value and the  optical depth is
approximated as the product of the real column density and the local opacity.
The lower two lines correspond to Equation \ref{cooling_approx.equ} evaluated according
to the polytropic method where the optical depth
is the product of the estimated column density and the estimated average 
opacity along the line of sight (see Appendix \ref{kappabar.sec}). In the case   
$\kappa \propto T^2$ this is $0.58$ times the local opacity. The bottom
line uses the full (star plus disc) potential in estimating the column
density whereas the curve above uses the disc potential only.

Recalling that in the optically thick regime the, cooling rate according
to Equation \ref{cooling_approx.equ}
scales as $\Sigma^{-2}$ and that the method over-estimates the column density
in disc geometries, it is no surprise that now the polytropic cooling method
yields a much smaller cooling rate, being at least an order of magnitude lower
in the case that only the disc potential is applied and more than
two orders of
magnitude if the total (star plus disc) potential is used. 

 When we compare each of these rates with the `true' cooling rate in the
radiative diffusion approximation
(dashed curve, Equation \ref{flux_diff.equ}) we find that the latter is now of comparable
magnitude to the cooling computed from Equation \ref{cooling_approx.equ}
using the true column density but is far larger than the estimates
from polytropic cooling. Thus in contrast to the case in spherical
geometry, there is no compensating effect that brings the polytropic estimate
into agreement with the `true' cooling rate.

\begin{figure}
\centering
\includegraphics[width=8.2cm]{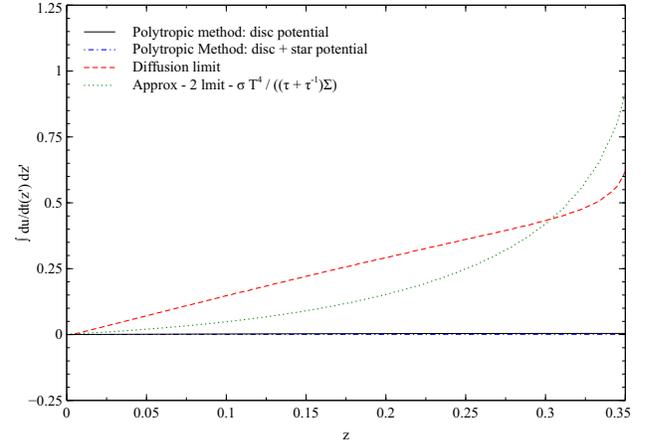}
\caption[]{Magnitude of the integrated cooling rate per unit area for the disc evaluated by the polytropic cooling method compared with the diffusion limit as well as Equation \ref{cooling_approx.equ} using analytic optical depths at radius $R=15$.} 
\label{disc_intcool.fig}
\end{figure}

  Figure \ref{disc_intcool.fig} illustrates how the specific cooling rates that are
compared in Figure \ref{disc_cool.fig} affect the total cooling rate per unit area. At each
value of $z$, the quantity plotted is given by the integral of the product
of specific
cooling rate and density from $z'=0$ to $z$. Note that in contrast to
Figure \ref{disc_cool.fig}, Figure \ref{disc_intcool.fig} is plotted on a linear scale. It illustrates that
at all heights, the use of Equation \ref{cooling_approx.equ} gives a reasonable (order
unity) estimate of the integrated cooling rate {\it provided one
has knowledge of the `true' column density}. Indeed, the total cooling
rate per unit area is within $25 \%$ of the result of using the radiative
diffusion approximation. The total cooling rate from the
polytropic method is, however, too low by a factor of at least $\sim 200$.
This emphasises that, in disc geometry, the most
pressing problem is to find a reliable estimator of the column density.
 
  We have also experimented with varying the opacity law in the disc geometry
and note that when the opacity is a less strongly increasing function
of temperature, there is a greater tendency (compared with the spherical case) 
for discs to be in the regime of net heating ({\it i.e.} the upper regions
receive a greater flux from the warm disc below than they radiate upwards --- the
outwardly increasing area element in spherical geometry makes this less
likely in the spherical case). Obviously, no local cooling prescription
can reproduce this effect. Thus any future efforts to find an improved
local cooling prescription in disc geometries will need to be carefully
calibrated in different opacity regimes.

\section{Conclusions}

  The polytropic method works quite well in spherical geometry. We have
shown that over a wide range of input parameters the mass integrated cooling
rate out to the $\tau = 1$ surface matches that from radiative
diffusion to within order unity or better, an error that is probably
negligible compared with other uncertainties such as those in the 
opacities. This good agreement is `fortuitous' in the sense that it
results from two nearly compensating discrepancies, but can nevertheless
be exploited in simulations in spherical symmetry.

 The  polytropic cooling estimate, however, performs poorly in optically
thick discs, regardless of whether one uses the potential from the disc
alone or that due to the star and the disc in order to estimate the column
density. This is due to the fact that the polytropic method over-estimates
the column density considerably (\S\ref{coldensity.sec}) since
it does not take account of the  lower column density route normal to the
disc plane. This translates into an {\it under-estimate} of the cooling
rate by between two and three orders of magnitude (Figures \ref{disc_cool.fig} and
\ref{disc_intcool.fig}), depending on whether
the stellar potential is included in the column density estimate.
We have shown however that (at least in the astrophysically
relevant ice opacity regime) the cooling can be well modelled in disc
geometries if only one can improve the column density estimate. This should clearly 
motivate further algorithmic developments in this area.
 
  The above is sufficient to say that the method needs to be used with
caution, though it does not necessarily negate results in the literature that 
use this method. For example, the outer regions of proto-stellar discs
may in fact be nearly isothermal at the back-ground temperature, in which
case all methods would agree on the thermal structure of the disc.
Alternatively, even if simulations enter regimes where the cooling rate
is strongly under-estimated, it may have a negligible effect on where,
for example, fragmentation occurs in the disc. This is because the
cooling rate in such discs is a very strong function of radius \citep{clarke-09}
and thus an under-estimate in cooling would cause only a modest increase
in the radius at which the disc fragments. 

  Simulators evidently need to balance these issues against the
  computational economies
offered by polytropic cooling. These results
can also inform those wishing to combine polytropic cooling with other
methods, for example the expedient of adding the polytropic cooling
method to flux limited diffusion (as in the `hybrid' method of
Forgan et al) relies on the former
being a minority contributor in the optically thick regions where
the latter is correct. We find that this requirement is amply fulfilled.
On the other hand, whether or not the hybrid is an improvement
over pure flux limited diffusion depends on the situation. Unlike 
the flux limited diffusion method, the hybrid method would yield the
correct cooling rate for non-irradiated optically thin gas. Neither method, however, is suitable for treating the optically thin regions above an
optically thick disc which are subject to irradiation from the disc below.
Whether or not this matters for the system dynamics obviously depends on
the topic at hand.

  Finally, 
we draw attention to the fact that an astrophysically relevant disc
structure (a marginally self-gravitating optically thick
steady state disc with opacity dominated by ice grains) 
can be modelled as an $n=2$ polytrope whose density and temperature
structure can be described analytically. The existence of such
an analytic solution may be useful in 
 testing future cooling
algorithms.
 
\section*{Acknowledgments}
We thank Ken Rice and Duncan Forgan for helpful discussions, Giuseppe Lodato for critical reading and Dimitris
Stamatellos, Anthony Whitworth and the referee for useful comments
on an earlier version of the paper. 

\bibliographystyle{mnras}
\bibliography{pt3paper}

\begin{thebibliography}{}

\bibitem[\protect\citeauthoryear{{Bate}}{{Bate}}{2009}]{bate}
{Bate} M.~R., 2009, \mnras, 392, 1363

\bibitem[\protect\citeauthoryear{{Bertin} \& {Lodato}}{{Bertin} \&
  {Lodato}}{1999}]{bertin_lodato}
{Bertin} G.,  {Lodato} G., 1999, \aap, 350, 694

\bibitem[\protect\citeauthoryear{{Bodenheimer} et~al.}{{Bodenheimer}
  et~al.}{1990}]{bodenheimer}
{Bodenheimer} P., {Yorke} H.~W., {Rozyczka} M.,  {Tohline} J.~E., 1990, \apj,
  355, 651

\bibitem[\protect\citeauthoryear{{Boley} et~al.}{{Boley} et~al.}{2007}]{boley}
{Boley} A.~C., {Durisen} R.~H., {Nordlund} {\AA}.,  {Lord} J., 2007, \apj, 665,
  1254

\bibitem[\protect\citeauthoryear{{Boss}}{{Boss}}{2008}]{boss-08}
{Boss} A.~P., 2008, \apj, 677, 607

\bibitem[\protect\citeauthoryear{{Boss} \& {Bodenheimer}}{{Boss} \&
  {Bodenheimer}}{1979}]{boss_bodenheimer-79}
{Boss} A.~P.,  {Bodenheimer} P., 1979, \apj, 234, 289

\bibitem[\protect\citeauthoryear{{Cai} et~al.}{{Cai} et~al.}{2008}]{cai}
{Cai} K., {Durisen} R.~H., {Boley} A.~C., {Pickett} M.~K.,  {Mej{\'{\i}}a}
  A.~C., 2008, \apj, 673, 1138

\bibitem[\protect\citeauthoryear{{Clarke}}{{Clarke}}{2009}]{clarke-09}
{Clarke} C.~J., 2009, \mnras, 396, 1066

\bibitem[\protect\citeauthoryear{{Forgan} et~al.}{{Forgan}
  et~al.}{2009}]{forgan}
{Forgan} D., {Rice} K., {Stamatellos} D.,  {Whitworth} A., 2009, \mnras, 394,
  882

\bibitem[\protect\citeauthoryear{{Gammie}, {Goodman}, \& {Ogilvie}}{{Gammie}
  et~al.}{2000}]{gammie}
{Gammie} C.~F., {Goodman} J.,  {Ogilvie} G.~I., 2000, \mnras, 318, 1005

\bibitem[\protect\citeauthoryear{{Hubeny}}{{Hubeny}}{1990}]{hubeny}
{Hubeny} I., 1990, \apj, 351, 632

\bibitem[\protect\citeauthoryear{{Krumholz} et~al.}{{Krumholz}
  et~al.}{2009}]{krumholz}
{Krumholz} M.~R., {Klein} R.~I., {McKee} C.~F., {Offner} S.~S.~R.,
  {Cunningham} A.~J., 2009, Science, 323, 754

\bibitem[\protect\citeauthoryear{{Larson}}{{Larson}}{2005}]{larson}
{Larson} R.~B., 2005, \mnras, 359, 211

\bibitem[\protect\citeauthoryear{{Masunaga} \& {Inutsuka}}{{Masunaga} \&
  {Inutsuka}}{2000}]{masunaga}
{Masunaga} H.,  {Inutsuka} S., 2000, \apj, 531, 350

\bibitem[\protect\citeauthoryear{{Mayer} et~al.}{{Mayer} et~al.}{2007}]{mayer}
{Mayer} L., {Lufkin} G., {Quinn} T.,  {Wadsley} J., 2007, \apjl, 661, L77

\bibitem[\protect\citeauthoryear{{Meru} \& {Bate}}{{Meru} \&
  {Bate}}{2011}]{meru-bate}
{Meru} F.,  {Bate} M.~R., 2011, \mnras, 411, L1

\bibitem[\protect\citeauthoryear{{Rice} et~al.}{{Rice} et~al.}{2003}]{rice}
{Rice} W.~K.~M., {Armitage} P.~J., {Bonnell} I.~A., {Bate} M.~R., {Jeffers}
  S.~V.,  {Vine} S.~G., 2003, \mnras, 346, L36

\bibitem[\protect\citeauthoryear{{Spiegel}}{{Spiegel}}{1957}]{spiegel}
{Spiegel} E.~A., 1957, \apj, 126, 202

\bibitem[\protect\citeauthoryear{{Stamatellos} \& {Whitworth}}{{Stamatellos} \&
  {Whitworth}}{2008}]{stamatellos_whitworth}
{Stamatellos} D.,  {Whitworth} A.~P., 2008, \aap, 480, 879

\bibitem[\protect\citeauthoryear{{Stamatellos} et~al.}{{Stamatellos}
  et~al.}{2007}]{stamatellos}
{Stamatellos} D., {Whitworth} A.~P., {Bisbas} T.,  {Goodwin} S., 2007, \aap,
  475, 37

\bibitem[\protect\citeauthoryear{{Toomre}}{{Toomre}}{1964}]{toomre}
{Toomre} A., 1964, \apj, 139, 1217

\bibitem[\protect\citeauthoryear{{Whitehouse} \& {Bate}}{{Whitehouse} \&
  {Bate}}{2004}]{wb04}
{Whitehouse} S.~C.,  {Bate} M.~R., 2004, \mnras, 353, 1078

\bibitem[\protect\citeauthoryear{{Whitehouse} \& {Bate}}{{Whitehouse} \&
  {Bate}}{2006}]{wb06}
{Whitehouse} S.~C.,  {Bate} M.~R., 2006, \mnras, 367, 32

\bibitem[\protect\citeauthoryear{{Whitehouse}, {Bate}, \&
  {Monaghan}}{{Whitehouse} et~al.}{2005}]{whitehouse}
{Whitehouse} S.~C., {Bate} M.~R.,  {Monaghan} J.~J., 2005, \mnras, 364, 1367

\bibitem[\protect\citeauthoryear{{Wolfire} \& {Cassinelli}}{{Wolfire} \&
  {Cassinelli}}{1987}]{wolfire}
{Wolfire} M.~G.,  {Cassinelli} J.~P., 1987, \apj, 319, 850

\bibitem[\protect\citeauthoryear{{Yorke} \& {Sonnhalter}}{{Yorke} \&
  {Sonnhalter}}{2002}]{yorke}
{Yorke} H.~W.,  {Sonnhalter} C., 2002, \apj, 569, 846

\end{thebibliography}

\appendix
\section{The Polytropic Cooling  Method}

The radiative cooling of an element within a cloud is determined by the optical depth of that point, that is the integrated opacity along the line-of-sight out of the cloud. The polytropic cooling method of \citet{stamatellos} aims to evaluate the radiative cooling of an element by calculating an approximate optical depth using only local variables at the point of interest.

In order to achieve this, the particle of interest is taken to be embedded at an arbitrary position in a spherically symmetric pseudo-cloud, modelled by a polytrope of index $n$. The density ($\rho$), potential ($\Psi$) and temperature ($T$) of the polytrope are given by the solutions of the Lane Emden equation, $\theta_n$,
\begin{eqnarray}
	\label{density.equ}
	\rho(\xi) &=& \rho_c\theta^n(\xi) \\
	\label{potential.equ}
	\Psi(\xi) &=& -4\pi G R_0^2\rho_c \phi(\xi) \\
	\label{tempscale.equ}	
	T(\xi) &=& T_c \theta(\xi)
\end{eqnarray}
where $ \phi(\xi) = -\xi_B\left(\frac{d\theta}{d\xi'}\right)_{\xi' = \xi_B} + \theta(\xi)$.

The parameters of the polytrope are set to reproduce the local conditions at the location of the particle. The column density, optical depth and radiative cooling are then calculated for the particle embedded in the pseudo-cloud. The value of the column density to the particle of interest is taken to be the average column density over all possible positions within the pseudo-cloud.

\subsection{Calibrating the Pseudo-Cloud}
Once the polytropic index of the barometric equation of state is set, the values of the central density, scale length and core temperature for the pseudo-cloud in which the particle of interested is embedded at a radius $r = R_0 \xi$ are calculated such that the known density ($\rho_i$), gravitational potential ($\Psi_i$) and temperature ($T_i$) at the position of the particle are reproduced.
\begin{eqnarray}
	\label{rhoc.equ}
	\rho_c &=& \rho_i \theta^{-n}(\xi) \\
	\label{R0.equ}
	R_0 &=& \sqrt{\frac{-\Psi_i \theta^n(\xi)}{4\pi G \rho_i \phi(\xi)}} \\
	\label{Tc.equ}
	T_c &=& T_i \theta^{-1}(\xi) 
\end{eqnarray}

\subsection{Column Density}
The column density (mass per unit area along a line-of-sight) from a particle at radius $r = \xi R_0$ along a radial line to the edge of the pseudo-cloud ($\xi = \xi_B$) is given by
\begin{equation}
	\Sigma = \int_r^{r_B} \rho(r) dr = R_0\int_\xi^{\xi_B} \rho(\xi') d\xi'
\end{equation}
Substituting for the density using Equation \ref{density.equ},
\begin{equation}
	\Sigma(\xi) = R_0\int_\xi^{\xi_B} \rho_c \theta^n(\xi') d\xi'
\end{equation}
And substituting for $\rho_c$ and $R_0$ using Equations \ref{rhoc.equ} and \ref{R0.equ},
\begin{equation}
	\Sigma(\xi) = \sqrt{\frac{-\Psi_i \rho_i}{4\pi G \phi(\xi) \theta^n(\xi)}} \int_\xi^{\xi_B} \theta^n(\xi') d\xi'
\end{equation}

When calculating the mean column density over all positions of the particle within the pseudo-cloud, the column density at a given radius $\Sigma(\xi)$ is weighted by the mass contained in a spherical shell radius $\xi$ thickness $d\xi$, $M(\xi)$ as this gives the probability of the particle being within that shell. The total mass of the cloud, $M$ is used to normalise the distribution.
\begin{equation}
	\bar\Sigma = \frac{1}{M}\int_0^{\xi_B} \Sigma(\xi)M(\xi) d\xi
\end{equation}
The mass enclosed in a spherical shell is given by the volume of the shell multiplied by the density, from the Lane-Emden function, $M(\xi) = 4\pi R_0^3 \rho_c \xi^2 \theta^n(\xi)$. The total mass of the cloud is $M = \int_0^{\xi_B} 4\pi R_0^3 \rho_c \xi^2 \theta^n(\xi)d\xi$.

Substituting these in to the integral gives
\begin{equation}
	\bar\Sigma = \left[ \int_0^{\xi_B} \xi^2 \theta^n(\xi)d\xi\right]^{-1} \int_0^{\xi_B} \xi^2 \Sigma(\xi)\theta^n(\xi) d\xi
\end{equation}
Evaluating the first integral and substituting in the column density,
\begin{equation}	
	\bar\Sigma = \left[ -\xi_B^2 \left(\frac{d\theta}{d\xi}\right)_{\xi = \xi_B}\right]^{-1} \int_0^{\xi_B} \left[ \int_\xi^{\xi_B} \theta^n(\xi')d\xi'\right] \sqrt{\frac{-\Psi_i\rho_i\theta^n(\xi)}{4\pi G \phi(\xi)}} \xi^2 d\xi
\end{equation}
Which can be written
\begin{equation}
	\label{SigmaBar.equ}
	\bar\Sigma = \zeta_n \sqrt{\frac{-\Psi_i\rho_i}{4\pi G}}
\end{equation}
Where $\zeta_n$ is just a constant for a given polytropic index.
\[	\zeta_n = \left[ -\xi_B^2 \left(\frac{d\theta}{d\xi}\right)_{\xi = \xi_B}\right]^{-1} \int_0^{\xi_B} \left[ \int_\xi^{\xi_B} \theta^n(\xi')d\xi'\right] \sqrt{\frac{\theta^n(\xi)}{\phi(\xi)}} \xi^2 d\xi \]
Values of $\zeta_n$ can be evaluated numerically and are given in Table \ref{zeta.tab}. It can be seen that the results for the column density are fairly insensitive to the polytropic index chosen in the approximation.

\begin{table}
	\label{zeta.tab}
	\begin{center}
	\begin{tabular}{|c|c|}
		\hline
		$n$ & $\zeta_n$ \\
		\hline
		0.0 & 0.3796 \\
		1.0 & 0.3756 \\
		2.0 & 0.3682 \\
		3.0 & 0.3600 \\
		4.0 & 0.3500 \\
		5.0 & 0.3322 \\
		\hline	
	\end{tabular}
	\end{center}
	\caption{Values of $\zeta_n$ for different polytropic indices (evaluated numerically).}
\end{table}
This mean is used as the approximate value of the column density at the point in question and depends only on the local potential and density. In this formula, the local potential conveys the structure of the surrounding space (rather than performing the integral along the line-of-sight).

\subsection{Optical Depth}
\label{taubar.sec}
The optical depth along a ray path with absorption coefficient $\alpha_u$ is defined as $\tau_u = \int \alpha_u(r) dr$ with $\alpha_u = \kappa_u\rho$. This can be averaged over all frequencies in the black body spectrum using the Rosseland-mean opacity $\kappa_R(\rho,T)$.

The optical depth of a particle at radius $r = \xi R_0$ in the pseudo-cloud can be calculated,
\begin{equation}
	\tau(\xi) = \int_\xi^{\xi_B} \kappa_R(\rho(\xi'),T(\xi'))\rho(\xi') R_0 d\xi'
\end{equation}
Substituting for the density and temperature using Equations \ref{density.equ} and \ref{tempscale.equ},
\begin{equation}
	\tau(\xi) = \int_\xi^{\xi_B} \kappa_R(\rho_c\theta^n(\xi'),T_c\theta(\xi'))\rho_c\theta^n(\xi') R_0 d\xi'
\end{equation}
And substituting for $\rho_c$, $T_c$ and $R_0$ using Equations \ref{rhoc.equ}, \ref{R0.equ} and \ref{Tc.equ},
\begin{equation}
	\tau(\xi) = \sqrt{\frac{-\Psi \rho_i \theta^n(\xi)}{4\pi G \phi(\xi)}} \int_\xi^{\xi_B} \kappa_R\left(\rho_i \left[\frac{\theta(\xi')}{\theta(\xi)}\right]^n,T_i\left[\frac{\theta(\xi')}{\theta(\xi)}\right]\right) \left[\frac{\theta(\xi')}{\theta(\xi)}\right]^n d\xi'
\end{equation}

The mass-weighted pseudo-mean optical depth (averaging over all possible positions of the particle within the pseudo-cloud) can be obtained as for the column density.
\begin{eqnarray*}
	\bar\tau &=& \left[ -\xi_B \left(\frac{d\theta}{d\xi'}\right)_{\xi' = \xi_B} \right]^{-1} \int_0^{\xi_B} \xi^2 \tau(\xi) \theta^n(\xi) d\xi \\
	\bar\tau &=& \left[ -\xi_B \left(\frac{d\theta}{d\xi'}\right)_{\xi' = \xi_B} \right]^{-1} \sqrt{\frac{-\Psi_i\rho_i}{4\pi G}} \\ &\ & \int_0^{\xi_B} \left[\int_\xi^{\xi_B} \kappa_R\left(\rho_i \left[\frac{\theta(\xi')}{\theta(\xi)}\right]^n,T_i\left[\frac{\theta(\xi')}{\theta(\xi)}\right]\right) \theta^n(\xi') d\xi'\right] \\ &\ & \sqrt{\frac{\theta^n(\xi)}{\phi(\xi)}} \xi^2 d\xi
\end{eqnarray*}
\subsection{Pseudo-Mean Mass Opacity}
\label{kappabar.sec}
Once the column density and optical depth have been calculated, one can define an average opacity along the line-of-sight,
\begin{eqnarray*}
	\bar\kappa &=& \frac{\bar\tau}{\bar\Sigma} \\
	\bar\kappa(\rho,T) &=& \left[ -\zeta_n\xi_B \left(\frac{d\theta}{d\xi'}\right)_{\xi' = \xi_B} \right]^{-1} \\ &\ & \int_0^{\xi_B} \left[\int_\xi^{\xi_B} \kappa_R\left(\rho \left[\frac{\theta(\xi')}{\theta(\xi)}\right]^n,T\left[\frac{\theta(\xi')}{\theta(\xi)}\right]\right) \theta^n(\xi') d\xi'\right] \sqrt{\frac{\theta^n(\xi)}{\phi(\xi)}} \xi^2 d\xi
\end{eqnarray*}
$\bar\kappa$ is a function only of $\rho$ and $T$, the local density and temperature, and does not depend on particle positions, therefore it need not be evaluated on-the-fly during simulations but can be evaluated in advance and looked-up from a table during simulations. Multiplying by the pseudo-mean column density, $\bar\Sigma$, recovers the optical depth.

\label{lastpage}
\end{document}